\documentclass[prb,aps,a4paper]{revtex4}

\usepackage{amsmath}
\usepackage{graphicx}
\usepackage{latexsym}
\usepackage{amsfonts}
\usepackage{amssymb}

\newcommand{\be}{\begin{equation}}
\newcommand{\ee}{\end{equation}}
\newcommand{\beq}{\begin{eqnarray}}
\newcommand{\eeq}{\end{eqnarray}}

\newcommand{\Tr}{ {\rm Tr} \, }

\begin{document}

\def\bbe{\mbox{\boldmath $e$}}
\def\bbf{\mbox{\boldmath $f$}}    
\def\bg{\mbox{\boldmath $g$}}
\def\bh{\mbox{\boldmath $h$}}
\def\bj{\mbox{\boldmath $j$}}
\def\bq{\mbox{\boldmath $q$}}
\def\bp{\mbox{\boldmath $p$}}
\def\br{\mbox{\boldmath $r$}}    

\def\bone{\mbox{\boldmath $1$}}    

\def\dr{{\rm d}}

\def\tb{\bar{t}}
\def\zb{\bar{z}}

\def\bC{\mbox{\boldmath $C$}}
\def\bG{\mbox{\boldmath $G$}}
\def\bH{\mbox{\boldmath $H$}}
\def\bK{\mbox{\boldmath $K$}}
\def\bM{\mbox{\boldmath $M$}}
\def\bN{\mbox{\boldmath $N$}}
\def\bO{\mbox{\boldmath $O$}}
\def\bQ{\mbox{\boldmath $Q$}}
\def\bR{\mbox{\boldmath $R$}}
\def\bS{\mbox{\boldmath $S$}}
\def\bT{\mbox{\boldmath $T$}}
\def\bU{\mbox{\boldmath $U$}}
\def\bV{\mbox{\boldmath $V$}}
\def\bZ{\mbox{\boldmath $Z$}}

\def\bcalS{\mbox{\boldmath $\mathcal{S}$}}
\def\bcalG{\mbox{\boldmath $\mathcal{G}$}}
\def\bcalE{\mbox{\boldmath $\mathcal{E}$}}

\def\bgG{\mbox{\boldmath $\Gamma$}}
\def\bgL{\mbox{\boldmath $\Lambda$}}
\def\bgS{\mbox{\boldmath $\Sigma$}}

\def\a{\alpha}
\def\b{\beta}
\def\g{\gamma}
\def\G{\Gamma}
\def\d{\delta}
\def\D{\Delta}
\def\e{\epsilon}
\def\ve{\varepsilon}
\def\z{\zeta}
\def\h{\eta}
\def\th{\theta}
\def\k{\kappa}
\def\l{\lambda}
\def\L{\Lambda}
\def\m{\mu}
\def\n{\nu}
\def\x{\xi}
\def\X{\Xi}
\def\p{\pi}
\def\P{\Pi}
\def\r{\rho}
\def\s{\sigma}
\def\S{\Sigma}
\def\t{\tau}
\def\f{\phi}
\def\vf{\varphi}
\def\F{\Phi}
\def\c{\chi}
\def\w{\omega}
\def\W{\Omega}
\def\Q{\Psi}
\def\q{\psi}

\def\ua{\uparrow}
\def\da{\downarrow}
\def\de{\partial}
\def\inf{\infty}
\def\ra{\rightarrow}
\def\bra{\langle}
\def\ket{\rangle}
\def\grad{\mbox{\boldmath $\nabla$}}

\title{Time dependent transport phenomena}

\author {G. Stefanucci, S. Kurth, E.K.U. Gross}
\affiliation{Institut f\"ur Theoretische Physik, Freie Universit\"at Berlin, 
Arnimallee 14, D-14195 Berlin, Germany,
and European Theoretical Sepctroscopy Facility (ETSF)}

\author{A. Rubio}
\affiliation{Departamento de F\'{i}sica de Materiales, Facultad de Ciencias 
Qu\'{i}micas, UPV/EHU, Unidad de Materiales Centro Mixto CSIC-UPV/EHU 
and Donostia International  Physics Center (DIPC), San Sebasti\'{a}n, Spain,
and European Theoretical Sepctroscopy Facility (ETSF)}

\begin{abstract}

\end{abstract}

\maketitle

\section{Introduction}

\begin{figure}[htbp]
\begin{center}
\includegraphics*[scale=0.8]{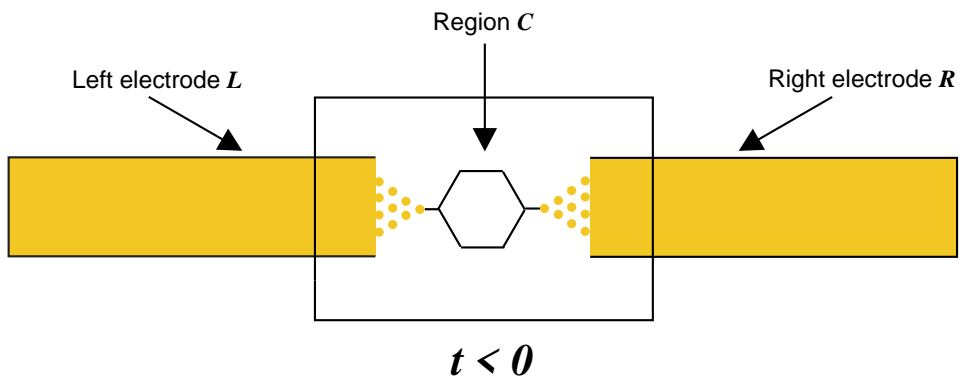}
\caption{Schematic sketch of the experimental setup described in the main text.  A
    central region which also includes few layers of the left and right
    electrodes is coupled to macroscopically large metallic reservoirs. The
    system is in equilibrium for negative times.}
\label{system}
\end{center}
\end{figure}

The aim of this review is to give a pedagogical introduction to our 
recently proposed {\em ab initio} theory of quantum transport. 
It is not intended to be a general overview of the field. For further 
information we refer the interested reader to 
Refs.~\onlinecite{Datta:95,HaugJauho:98,cufari}. 
The nomenclature {\em quantum transport} has been coined for the phenomenon of
electron motion through constrictions of transverse dimensions smaller than the
electron wavelength, e.g., quantum-point contacts, quantum wires,
molecules, etc. The typical experimental setup is displayed in Fig. 
\ref{system} where a central region $C$ of meso- or nano-scopic size is coupled to
two metallic electrodes $L$ and $R$ which play the role of charge reservoirs. The whole
system is initially (at time $t<0$) in a well defined equilibrium configuration, described by a
{\em unique} temperature and chemical potential (thermodynamic consistency).  
The charge density of the electrodes is perfectly balanced and no
current flows through the junction.

As originally proposed by Cini, \cite{cini} we may
drive the system out of equilibrium by exposing the electrons to an external
time-dependent potential which is local in time and space.  For instance, we may
switch on an electric field by putting the system between two capacitor plates
far away from the system boundaries. The dynamical
formation of dipole layers screens the potential drop along the electrodes and
the total potential turns out to be uniform in the left and right bulks.
Accordingly, the potential drop is entirely limited to the central region. As
the system size increases, the remote parts are less disturbed by the junction,
and the density inside the electrodes approaches the equilibrium bulk density.

The Cini scheme can be combined with Time Dependent Density Functional Theory 
(TDDFT).\cite{rg} In this theory, the time-dependent density of an interacting system 
moving in an external, time-dependent local potential can be calculated via a fictitious 
system of non-interacting electrons moving in a local, effective time-dependent potential. 
Therefore this theory is in principle well suited for the treatment of nonequilibrium transport 
problems.\cite{StefanucciAlmbladh:04} However, as far as the leads are treated as 
\textit{noninteracting}, it is not obvious that in the 
long-time limit a steady-state current can ever develop. The reason 
behind the uncertainty is that the bias 
represents a large perturbation and, in the absence of dissipative 
effects, e.g., electron-electron or electron-phonon scattering, 
the return of time-translational invariance is not granted. In this 
review we will show that the total current tends to a steady-state value 
provided the effective potential of TDDFT is independent of time and 
space in the left and right bulks. Also, the physical mechanism 
leading to the dynamical formation of a steady state is clarified. 

It should be mentioned that there has been already considerable activity 
in the density functional theory (DFT) community to describe transport 
phenomena through systems like the one in Fig. \ref{system}. 
Most approaches are limited to the steady-state regime and are based on a 
self-consistency procedure first proposed by Lang.\cite{lang} In this steady-state 
approach based on DFT, exchange and correlation is 
approximated by the static Kohn-Sham (KS) potential and the charge density is obtained 
self-consistently in the presence of the steady current.
However, the original justification involved subtle points such as different 
Fermi levels deep inside the left and right electrodes (which is not 
thermodynamically consistent) and the implicit reference of non-local perturbations 
such as tunneling Hamiltonians within a DFT framework. (For a 
detailed discussion we refer to Ref. \onlinecite{saprb04}.) 
Furthermore, the transmission functions computed from static DFT have resonances at the 
non-interacting KS excitation energies which in general do not 
coincide with the true excitation energies. 

Our TDDFT formulation, as opposed to the static DFT formulation, is 
thermodynamically consistent, is not limited to the steady-state 
regime (we can study transients, AC responses, etc.) and has the 
extra merit of accessing the true excitation energies of interacting 
systems.\cite{PetersilkaGossmannGross:96}

We will first use the nonequilibrium Green's function (NEGF) technique to 
discuss the implications of our approach. For those readers that are  
not familiar with the Keldysh formalism and with NEGF, in Section 
\ref{keldform} we give an elementary introduction to the Keldysh 
contour, the Keldysh-Green functions and the Keldysh book-keeping. 
The aim of this Section is to derive some of the identities needed for 
the discussion (thus providing a self-contained presentation) and to 
establish the basic notation. 
In Section \ref{negf} we set up the theoretical framework by 
combining TDDFT and NEGF. An exact expression for the time-dependent 
total current $I(t)$ is written in terms of Green functions projected in 
region $C$. It is also shown that a steady-state 
regime develops provided 1) the KS Hamiltonian {\em globally} converges to an 
asymptotic KS Hamiltonian when $t\ra\inf$, 2) the electrodes form a continuum of states 
(thermodynamic limit), and 3) the local density of states is a smooth function in the 
central region. It is worth noting that the steady-state current 
results from a pure dephasing mechanism in the fictitious KS system. 
Also, the resulting steady current only depends on the KS potential 
at $t=\inf$ and not on its history. However, the KS potential 
might depend on the history of the 
external applied potential and the resulting steady-state 
current might be history dependent.
A practical scheme to calculate $I(t)$ is presented in Section 
\ref{pract}. The main idea is to propagate the KS orbitals in region $C$ 
only, without dealing with the infinite and non-periodic 
system.\cite{ksarg2005} We 
first show how to obtain the KS eigenstates $\q_{s}$ of the undisturbed system 
in Section \ref{ks0}. Then, in Section \ref{kst} we describe an algorithm 
for propagating $\q_{s}$ under the influence of a time-dependent 
disturbance. The numerical approach of Section \ref{pract} is completely 
general and can be applied to any system having the geometry sketched 
in Fig. \ref{system}. In order to demonstrate the feasibility of the 
scheme we implement it for one-dimensional model systems in Section 
\ref{details}. Here we study the dynamical current response of several 
systems perturbed by DC and AC biases. We verify that 
for noninteracting electrons the steady-state current does not depend on 
the history of the applied bias. Also, we present preliminary 
results on net currents in unbiased systems as obtained by pumping 
mechanisms. We summarize our findings and draw our 
conclusions in Section \ref{conclusion}.

\section{The Keldysh formalism}
\label{keldform}

\subsection{The Keldysh contour}

In quantum mechanics we associate to any observable quantity 
$O$ a hermitian operator $\hat{O}$. The expectation  
$\bra\Q|\hat{O}|\Q\ket$ gives the value of $O$ when 
the system is described by the state $|\Q\ket$. For an isolated system 
the Hamiltonian $\hat{H}_{0}$ does not depend on time, and the expectation value of 
{\em any} observable quantity is constant provided $|\Q\ket$ is an 
eigenstate of $\hat{H}_{0}$. In this Section we discuss how to 
describe systems which are not isolated but perturbed by external 
fields. Without loss of generality, we assume that the system is isolated for negative 
times $t$ and that $\hat{H}(t<0)=\hat{H}_{0}$. The evolution of the 
state $|\Q\ket$ is governed by the Schr\"odinger equation
$i\frac{\dr}{\dr t}|\Q(t)\ket=\hat{H}(t)|\Q(t)\ket$, 
and, correspondingly, the value of $O$ evolves in time as 
$O(t)=\bra\Q(t)|\hat{O}|\Q(t)\ket$. The time-evolved state 
$|\Q(t)\ket=\hat{S}(t;0)|\Q(0)\ket$, where the evolution operator 
$\hat{S}(t;t')$ can be formally written as 
\begin{equation}
\hat{S}(t;t')=\left\{
\begin{array}{ll}
    T\,{\rm e}^{-i\int_{t'}^{t}\dr\bar{t}\,\hat{H}(\bar{t})} &
    \quad t>t' \\
    \overline{T}\,{\rm e}^{-i\int_{t'}^{t}\dr\bar{t}\,\hat{H}(\bar{t})} &
    \quad t<t'
\end{array}    
\right..
\label{eo}
\end{equation}
In Eq. (\ref{eo}), $T$ is the time-ordering operator and 
rearranges the operators in chronological order with later times to 
the left; $\overline{T}$ is the anti-chronological time-ordering operator.  
The evolution operator is unitary and satisfies the group property 
$\hat{S}(t;t_{1})\hat{S}(t_{1};t')=\hat{S}(t;t')$ for any $t_{1}$. 
It follows that $O(t)$ is the average on the initial 
state $|\Q(0)\ket$ of the operator $\hat{O}$ in the Heisenberg representation,  
$\hat{O}_{H}(t)=\hat{S}(0;t)\hat{O}\hat{S}(t;0)$, i.e.,
\begin{equation}
O(t)=\bra\Q(0)|\hat{S}(0;t)\hat{O}\hat{S}(t;0)|\Q(0)\ket
=\bra\Q(0)|\overline{T}{\rm e}^{-i\int_{t}^{0}\dr\bar{t}\,\hat{H}(\bar{t})}
\;\hat{O}\;T{\rm e}^{-i\int_{0}^{t}\dr\bar{t}\,\hat{H}(\bar{t})}|\Q(0)\ket.
\label{evba}
\end{equation}
\begin{figure}[htbp]
\begin{center}
\includegraphics*[scale=0.9]{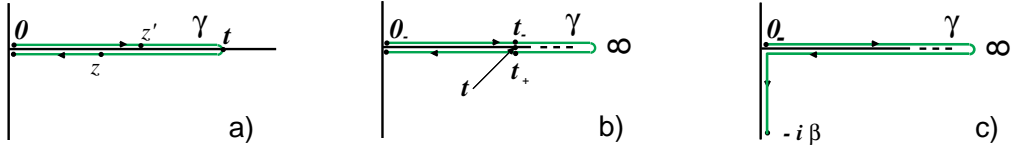}
\caption{a) The oriented contour $\g$ described in the main 
text with a forward and a backward branch between $0$ and $t$. 
According with the orientation the point $z$ is later than the point 
$z'$. b) The extended oriented contour $\g$ described in the main 
text with a forward and a backward branch between $0$ and $\inf$.
For any physical time $t$ we have two points $t_{\pm}$ on $\g$ at the 
same distance from the origin. c) The generalization of the original Keldysh 
contour. A vertical track going from $0$ to $-i\b$ has been added and, 
according with the orientation chosen, any point lying on it is later than a 
point lying on the forward or backward branch.}
\label{ksv}
\end{center}
\end{figure}

We can now design an oriented contour $\g$ with a forward branch going 
from $t=0$ to $t$ and a backward branch coming back from $t$ 
and ending in $t=0$, see Fig. \ref{ksv}.a. Denoting with $\zb$ the variable 
running on $\g$, Eq. (\ref{evba}) can be formally recast as follows
\begin{equation}
O(t)=\bra\Q(0)|T_{\rm K}\left\{{\rm e}^{-i\int_{\g}\dr 
\zb\,\hat{H}(\zb)}
\;\hat{O}(t)\right\}|\Q(0)\ket.
\label{ev3}
\end{equation}
The contour ordering operator $T_{\rm K}$ moves the operators 
with ``later'' contour variable to the left. A point $z$ is later than 
a point $z'$ if $z'$ is closer to the starting point, see Fig. 
\ref{ksv}.a.   
In Eq. (\ref{ev3}), $\hat{O}(t)$ is {\em not} the operator in the 
Heisenberg representation [the latter is denoted with $\hat{O}_{H}(t)$]. Actually, 
$\hat{O}(t)=\hat{O}$ for any $t$. 
The reason of the time argument stems from the need of specifying 
the position of the operator $\hat{O}$ in the contour ordering.

Let us now extend the contour $\g$ up to infinity, as shown in Fig. \ref{ksv}.b. 
For any physical time $t$ there are two points $z=t_{+}$ and $z=t_{-}$ on $\g$; 
$t_{-}$ lies on the forward branch while $t_{+}$ lies on the 
backward branch and it is later than $t_{-}$ according with the 
orientation chosen. We have
$
T_{\rm K}\{{\rm e}^{-i\int_{\g}\dr 
\zb\,\hat{H}(\zb)}
\;\hat{O}(t_{-})\}
=\hat{S}(0;\inf)\hat{S}(\inf;t)\hat{O}(t)\hat{S}(t;0)
=\hat{S}(0;t)\hat{O}\hat{S}(t;0),
$
and similarly
$
T_{\rm K}\{{\rm e}^{-i\int_{\g}\dr 
\zb\,\hat{H}(\zb)}
\;\hat{O}(t_{+})\}=
\hat{S}(0;t)\hat{O}(t)\hat{S}(t;\inf)\hat{S}(\inf;0)
=\hat{S}(0;t)\hat{O}\hat{S}(t;0).
$
Thus, the expectation value $O(t)$ in Eq. (\ref{ev3}) is also given by 
the formula
\begin{equation}
O(z)=\bra\Q(0)|T_{\rm K}\left\{{\rm e}^{-i\int_{\g}\dr 
\zb\,\hat{H}(\zb)}
\;\hat{O}(z)\right\}|\Q(0)\ket.
\label{ev5}
\end{equation}
where $\g$ is the contour in Fig. \ref{ksv}.b; $\g$ is called 
the {\em Keldysh contour}.\cite{kb-book,keldysh-jetp1965} 
In Eq. (\ref{ev5}) the variable $z$ can be either $t_{-}$ or $t_{+}$ 
and $O(t_{-})=O(t_{+})=O(t)$.

The Keldysh contour can be further extended to account for statistical 
averages.\cite{daniele} In statistical physics a system is described by the 
density matrix $\hat{\r}=\sum_{n}w_{n}|\Q_{n}\ket\bra\Q_{n}|$ with 
$w_{n}$ the probability of finding the system in the state 
$|\Q_{n}\ket$ and $\sum_{n}w_{n}=1$. The states $|\Q_{n}\ket$ may not 
be orthogonal. We say that the system is in a pure state if 
$\hat{\r}=|\Q\ket\bra\Q|$. In a system described by a density matrix 
$\hat{\r}(0)$ at $t=0$, the time-dependent value of the observable $O$ 
is a generalization of Eq. (\ref{ev5}), i.e.,
$
O(z)=\sum_{n}w_{n}\bra\Q_{n}(0)|T_{\rm K}\{{\rm e}^{-i\int_{\g}\dr 
\zb\,\hat{H}(\zb)}
\;\hat{O}(z)\}|\Q_{n}(0)\ket.
$
Among all possible density matrices there is one 
that is very common in physics and describes a system in thermal 
equilibrium:
$
\hat{\r}=\exp[-\b (\hat{H}_{0}-\m \hat{N})]/\mathcal{Z}, 
$
with the inverse temperature $\b$, the chemical potential $\m$,  
the operator $\hat{N}$ corresponding to the total number of particles 
and the grand-partition function $\mathcal{Z}=\Tr\exp[-\b (\hat{H}_{0}-\m\hat{N})]$. 
Assuming that $\hat{H}_{0}$ and $\hat{N}$ 
commute, the statistical average $O(z)$ with the thermal density 
matrix can be written as
$
O(z)=\Tr[\,{\rm e}^{\b\m\hat{N}}{\rm e}^{-\b\hat{H}_{0}}
T_{\rm K}\{{\rm e}^{-i\int_{\g}\dr \zb\,\hat{H}(\zb)}
\;\hat{O}(z)\}\,]/\mathcal{Z}.
$
We can now extend further the Keldysh contour as shown in Fig. \ref{ksv}.c and define 
$\hat{H}(z)=\hat{H}_{0}$ for any $z$ on the vertical track. With this 
definition $\hat{H}(z)$ is continuous along the entire contour since 
$\hat{H}(0)=\hat{H}_{0}$. According to the orientation displayed in 
the figure, any point lying on 
the vertical track is later than a point lying on the forward or backward 
branch. We use this observation to rewrite $O(z)$ as 
\begin{equation}
O(z)=\frac{\Tr\left[ {\rm e}^{\b\m\hat{N}} 
T_{\rm K}\left\{{\rm e}^{-i\int_{\g}\dr 
\zb\,\hat{H}(\zb)}
\;\hat{O}(z)\right\}\right]}{\Tr\left[{\rm e}^{\b\m\hat{N}} 
T_{\rm K}\left\{{\rm e}^{-i\int_{\g}\dr 
\zb\,\hat{H}(\zb)}\right\}\right]},
\label{ev8}
\end{equation}   
where $T_{\rm K}$ is now the ordering operator on the extended contour. 
It is worth noting that the denominator in the above expression is simply 
$\mathcal{Z}$. 
We have already shown that choosing $z$ on one of the two horizontal 
branches, Eq. (\ref{ev8}) yields the time-dependent statistical average 
of the observable $O$. On the other hand, if $z$ lies on 
the vertical track 
$
O(z)=\Tr[\,{\rm e}^{\b\m\hat{N}}{\rm e}^{-i\int_{z}^{-i\b}\hat{H}_{0}}
\hat{O}{\rm e}^{-i\int_{0}^{z}\hat{H}_{0}}\,]/\mathcal{Z}=
\Tr[\,{\rm e}^{-\b(\hat{H}_{0}-\m\hat{N})}
\hat{O}]/\mathcal{Z},
$
where the cyclic property of the trace has been used. The result
is independent of $z$ and coincides with the thermal average of 
the observable $O$. 

To summarize, in Eq. (\ref{ev8}) the 
variable $z$ lies on the contour of Fig. \ref{ksv}.c; the r.h.s. gives 
the time-dependent statistical average of the observable quantity $O$ 
when $z$ lies on the forward or backward branch, and the statistical average 
before the system is disturbed when $z$ lies on the vertical track. 

\subsection{The Keldysh-Green function}

The idea presented in the previous Section can be used to define correlators 
of many operators on the extended Keldysh contour. The Keldysh-Green 
function $\bG$ is the correlator of two field operators $\q(\br)$ and $\q^{\dag}(\br)$ 
which obey the anticommutation relations $\{\q(\br),\q^{\dag}(\br')\}=\d(\br-\br')$. 
It is defined as 
\be
\bra\br|\bG(z;z')|\br'\ket=\frac{1}{i}
\frac{\Tr\left[ {\rm e}^{\b\m\hat{N}} 
T_{\rm K}\left\{{\rm e}^{-i\int_{\g}\dr 
\zb\,\hat{H}(\zb)}
\;\q(\br,z)\q^{\dag}(\br',z')\right\}\right]}
{\Tr\left[{\rm e}^{\b\m\hat{N}} T_{\rm K}\left\{{\rm e}^{-i\int_{\g}\dr 
\zb\,\hat{H}(\zb)}\right\}\right]},
\ee
where the contour variable in the field operators specifies the position 
in the contour ordering (there is no true dependence on $z$ in $\q$ 
and $\q^{\dag}$). Here and in the following we use boldface to indicate 
matrices in one-electron labels, {\em e.g.}, $\bG$ is a matrix and 
$\bra\br|\bG|\br'\ket$ is the $(\br,\br')$ matrix element of $\bG$. 
Due to the contour ordering operator $T_{\rm K}$, the Green 
function $\bG$ has the following structure
\begin{equation}
\bG(z;z')=\theta(z,z')\bG^{>}(z;z')+\theta(z',z)\bG^{<}(z;z'),
\label{kgf}
\end{equation}
where $\th(z,z')=1$ if $z$ is later than $z'$ on the 
contour and zero otherwise. We say that a two-point function 
on the contour having the above structure belongs to the {\em Keldysh space}. 
The Green function $\bG(z;z')$ obeys an important cyclic relation 
on the extended Keldysh contour. As we shall see, the relations 
below play a crucial role since they provide the boundary conditions 
for solving the Dyson equation. Choosing 
$z=0_{-}$ we find 
\begin{equation}
\bra\br|\bG(0_{-};z')|\br'\ket=    
-\frac{1}{i}\frac{\Tr\left[{\rm e}^{\b\m\hat{N}}
T_{\rm K}\left\{{\rm e}^{-i\int_{\g}\dr 
\zb\,\hat{H}(\zb)}
\q^{\dag}(\br',z')\right\}\q(\br)\right]}
{\Tr\left[{\rm e}^{\b\m\hat{N}} 
T_{\rm K}\left\{{\rm e}^{-i\int_{\g}\dr 
\zb\,\hat{H}(\zb)}\right\}\right]},
\label{bc1}
\end{equation}
where we have taken into account that $0_{-}$ is the earliest time 
and therefore $\q(\br,0_{-})$ is always moved to the right when acted 
upon by $T_{\rm K}$. The extra 
minus sign in the r.h.s. comes from the contour ordering. More 
generally, rearranging the field operators $\q$ and $\q^{\dag}$ (later 
arguments to the left), we also have to multiply by $(-1)^{P}$, 
where $P$ is the parity of the permutation.
Inside the trace we can move $\q(\br)$ to the left. Furthermore, we 
can exchange the position of $\q(\br)$ and ${\rm e}^{\b\m\hat{N}}$ by 
noting that $\q(\br){\rm e}^{\b\m\hat{N}}={\rm e}^{\b\m(\hat{N}+1)}\q(\br)$. 
Using the fact that $T_{\rm K}$ moves operators with later times to the 
left we have $\q(\br)T_{\rm K}\{\ldots\}=T_{\rm 
K}\{\q(\br,-i\b)\ldots\}$. Therefore, we conclude that
\begin{equation}
\bG(0_{-};z')=-{\rm e}^{\b\m}\bG(-i\b;z'), \quad\quad
\bG(z;0_{-})=-{\rm e}^{-\b\m}\bG(z;-i\b),
\label{bclr}
\end{equation}
where the second of these relations can be obtained in a similar way.
Eq. (\ref{bclr}) are the so called Kubo-Martin-Schwinger 
(KMS) boundary conditions.\cite{kubo,martin} 

\subsection{The Keldysh book-keeping}
\label{kbk}

In this Section we derive some of the identities that we will use  
for dealing with time-dependent transport phenomena. A systematic and more exhaustive 
discussion can be found in Ref. \onlinecite{vldsavb2005}. 

Let $k(z;z')$ belong to the Keldysh space: $k(z;z')=\th(z,z')k^{>}(z;z')+
\th(z',z)k^{<}(z;z')$.  For any $k(z;z')$ in the Keldysh 
space we define the {\em greater} and {\em lesser} 
functions on the physical time axis
\be
k^{>}(t;t')\equiv k(t_{+};t'_{-}),\quad
k^{<}(t,t')\equiv k(t_{-};t'_{+}).
\label{k><}
\ee
We also define the {\em left} and {\em right} functions with one argument 
$t$ on the physical time axis and the other $\t$ on the vertical track
\begin{equation}
k^{\rceil}(t;\t)\equiv k(t_{\pm};\t),\quad
k^{\lceil}(\t,t)\equiv k(\t;t_{\pm}).
\label{kceil}
\end{equation}
In the definition of $k^{\rceil}$ and $k^{\lceil}$ we can arbitrarily 
choose $t_{+}$ or $t_{-}$ since $\t$ is later than both of them. The 
symbols ``$\rceil$'' and ``$\lceil$'' have been chosen in order to 
help the visualization of the time arguments. For instance, 
``$\rceil$'' has a horizontal segment followed by a vertical one;  
correspondingly, $k^{\rceil}$ has a first argument which is real 
(and thus lies on the horizontal axis) and a second argument which is 
imaginary (and thus lies on the vertical axis). We will also use the 
convention of denoting with Latin letters the real time and with 
Greek letters the imaginary time.

It is straightforward to show that if $a(z;z')$ and $b(z;z')$ belong to the 
Keldysh space, then $c(z;z')=\int_{\g}\dr \zb \;a(z;\zb)b(\zb;z')$
also belongs to the Keldysh space. From the definitions 
(\ref{k><}-\ref{kceil}) we find
\begin{eqnarray}
c^{>}(t;t')&=&
\int_{0_{-}}^{t'_{-}}\dr\zb\;a^{>}(t_{+};\zb)b^{<}(\zb;t'_{-})+
\int_{t'_{-}}^{t_{+}}\dr\zb\;a^{>}(t_{+};\zb)b^{>}(\zb;t'_{-})+
\int_{t_{+}}^{-i\b}\dr\zb\;a^{<}(t_{+};\zb)b^{>}(\zb;t'_{-})=
\nonumber \\ &=&
\int_{0}^{t'}\dr\bar{t} \;a^{>}(t;\bar{t})b^{<}(\bar{t};t')+
\int_{t'}^{t}\dr\bar{t}\;a^{>}(t;\bar{t})b^{>}(\bar{t};t')+
\int_{t}^{0}\dr\bar{t}\;a^{<}(t;\bar{t})b^{>}(\bar{t};t')+
\int_{0}^{-i\b}\dr\bar{\t}\;a^{\rceil}(t;\bar{\t})b^{\lceil}(\bar{\t};t').
\label{cgc}
\end{eqnarray}
The second integral on the r.h.s. is an ordinary integral on the 
real axis of two well defined functions and may be rewritten as
$\int_{t'}^{t}\dr\bar{t}\;a^{>}(t;\bar{t})b^{>}(\bar{t};t')=
\int_{t'}^{0}\dr\bar{t}\;a^{>}(t;\bar{t})b^{>}(\bar{t};t')
+\int_{0}^{t}\dr\bar{t}\;a^{>}(t;\bar{t})b^{>}(\bar{t};t')$. 
Using this relation, Eq. (\ref{cgc}) becomes
\begin{equation}
c^{>}(t;t')=\int_{0}^{\inf}\dr\bar{t} \;[a^{>}(t;\bar{t})b^{\rm A}(\bar{t};t')
+a^{\rm R}(t;\bar{t})b^{>}(\bar{t};t')]+
\int_{0}^{-i\b}\dr\bar{\t}\;a^{\rceil}(t;\bar{\t})b^{\lceil}(\bar{\t};t'),
\label{c>}
\end{equation}
where we have introduced two other functions on the physical time axis 
\begin{equation}
k^{\rm R}(t;t')\equiv \theta(t-t')[k^{>}(t;t')-k^{<}(t;t')],
\;
k^{\rm A}(t;t')\equiv -\theta(t'-t)[k^{>}(t;t')-k^{<}(t;t')].
\label{kra} 
\end{equation}
The {\em retarded} function $k^{\rm R}(t;t')$ vanishes for $t<t'$, while 
the {\em advanced} function $k^{\rm A}(t;t')$ vanishes for $t>t'$. 
A relation similar to Eq. (\ref{c>}) can be obtained for the lesser 
component $c^{<}$.
It is convenient to introduce a short hand notation for integrals 
along the physical time axis and for those between 0 and $-i\b$. The 
symbol ``$\cdot$'' will be used to write 
$\int_{0}^{\inf}\dr\tb f(\tb)g(\tb)$ as $f\cdot g$, while 
the symbol ``$\star$'' will be used to write 
$\int_{0}^{-i\b}\dr\bar{\t}f(\bar{\t})g(\bar{\t})$ as $f\star g$. Then
\begin{equation}
c^{>}=a^{>}\cdot b^{\rm A}+a^{\rm R}\cdot b^{>}+a^{\rceil}\star 
b^{\lceil},
\quad\quad
c^{<}=a^{<}\cdot b^{\rm A}+a^{\rm R}\cdot b^{<}+a^{\rceil}\star 
b^{\lceil}.
\label{c><}
\end{equation}
Eq. (\ref{c><}) can be used to extract the retarded 
and advanced component of $c$. By definition
$c^{\rm R}(t;t')=\theta(t-t')[c^{>}(t;t')-c^{<}(t;t')]$ and therefore
\be
c^{\rm R}(t;t')=\th(t-t')\int_{0}^{\inf}\dr\bar{t}\,
a^{\rm R}(t;\bar{t})[b^{>}(\bar{t};t')- b^{<}(\bar{t};t')]
+\th(t-t')\int_{0}^{\inf}\dr\bar{t}\,
[a^{>}(t;\bar{t})-a^{<}(t;\bar{t})]b^{\rm A}(\bar{t};t').
\ee
Due to the $\th$-function, we have $t>t'$ for $c^{\rm R}\neq 0$. 
In the second term on the r.h.s. 
$b^{\rm A}(\bar{t};t')$ contains a $\th(t'-\bar{t})$ and hence it must be $t>\bar{t}$; 
therefore we can replace the difference in the square 
bracket with $a^{\rm R}$. Then we break the first term on the r.h.s. 
in two pieces by inserting $\th$-functions: one for $\bar{t}<t'$ and the other 
for $\bar{t}>t'$. In compact notation we end up with
\begin{equation}
c^{\rm R}=a^{\rm R}\cdot b^{\rm R},
\quad\quad
c^{\rm A}=a^{\rm A}\cdot b^{\rm A},
\label{cra}
\end{equation}
where the second relation can be proven in a similar way.
It is worth noting that in the expressions for $c^{\rm R}$ and $c^{\rm A}$ 
no integration along the imaginary track is required. For later 
purposes we also define  the {\em Matsubara} function $k^{\rm M}(\t;\t')$ 
with both the arguments in the interval $(0,-i\b)$:
\be
k^{\rm M}(\t;\t')\equiv k(z=\t;z'=\t').
\label{km}
\ee
It is straightforward to prove the following identities 
\begin{equation}
c^{\rceil}=a^{\rm R}\cdot b^{\rceil}+a^{\rceil}\star b^{\rm M},
\quad\quad
c^{\lceil}=a^{\lceil}\cdot b^{\rm A}+a^{\rm M}\star b^{\lceil},
\quad\quad
c^{\rm M}=a^{\rm M}\star b^{\rm M}.
\label{cm}
\end{equation}

Finally, we consider the case of a Keldysh function 
$k(z;z')$ multiplied on the left by a scalar function $l(z)$.
The function $k_{l}(z;z')=l(z)k(z;z')=\theta(z,z')l(z)k^{>}(z;z')+
\theta(z';z)l(z)k^{<}(z;z')$ 
and hence belongs to the Keldysh space. The Keldysh components can 
be extracted using the definitions 
(\ref{k><},\ref{kceil},\ref{kra},\ref{km}). Choosing 
for instance $z=t_{+}$ and $z'=t'_{-}$ we find 
$k^{>}_{l}(t;t')=l(t)k^{>}(t;t')$ and similarly for $z=t_{-}$ and $z'=t'_{+}$ we find 
$k^{<}_{l}(t;t')=l(t)k^{<}(t;t')$. More generally, the function $l$ 
is simply a prefactor: $k^{\rm x}_{l}=lk^{\rm x}$, where $\rm x$ 
is one of the Keldysh components ($\lessgtr$, ${\rm 
R,\;A}$, $\rceil,\;\lceil$, ${\rm M}$). The same 
is true for $k_{r}(z;z')=k(z;z')r(z')$, where $r(z')$ is a scalar 
function: $k^{\rm x}_{r}=k^{\rm x}r$.

\section{Quantum transport using TDDFT and NEGF}
\label{negf}

\subsection{Merging the Keldysh and TDDFT formalisms}

The one-particle scheme of TDDFT corresponds to a fictitious system of 
noninteracting electrons described by the Kohn-Sham (KS) Hamiltonian 
$\hat{H}_{s}(z)=\int \dr \br \dr\br' \q^{\dag}(\br)\bra \br|\bH_{s}(z)
|\br'\ket\q(\br')$. 
The potential $v_{s}({\bf r},t)$ experienced by the electrons in 
the free-electron Hamiltonian $\mbox{\boldmath $H$}_{s}(t)$ 
is called the KS potential and it is given by the sum of the external 
potential, the Coulomb potential of the nuclei, the Hartree 
potential and the exchange-correlation potential $v_{\rm xc}$. 
The latter accounts for the complicated many-body effects and is 
obtained from an exchange-correlation action functional, 
$v_{\rm xc}({\bf r},t)=\delta A_{\rm xc}[n]/\delta n({\bf r},t)$  
(as pointed out in Ref. \onlinecite{vanleeuwen}, the causality 
and symmetry properties require that the action functional 
$A_{\rm xc}[n]$ is defined on the Keldysh contour). 
$A_{\rm xc}$ is a functional of the density and of the 
initial density matrix. In our case, the initial density matrix 
is the thermal density matrix  
which, due to the extension of the Hohenberg-Kohn theorem\cite{hk1964}
to finite temperatures,\cite{m1965} also 
is a functional of the density. 
We should mention here that an alternative formulation based on TDDFT has 
been recently proposed by Di Ventra and Todorov\cite{dvt2004}. In 
their approach the system is initially unbalanced and therefore the 
exchange-correlation functional depends on the initial state and not 
only on the density.

The fictitious Keldysh-Green function $\bcalG(z;z')$ of the KS 
system satisfies a one-particle equation of motion 
\begin{equation}
\left\{i\frac{\overrightarrow{\dr}}{\dr z}\bone-\bH_{s}(z)\right\}
\bcalG(z;z')=\bone\d(z-z'),
\nonumber
\end{equation}
\begin{equation}
\bcalG(z;z')\left\{-i\frac{\overleftarrow{\dr}}{\dr z'}\bone-\bH_{s}(z')\right\}=
\bone\d(z-z'),
\label{eomg}
\end{equation}
with KMS boundary conditions (\ref{bclr}). 
In Eqs. (\ref{eomg}) the arrow specifies where the derivative along the contour 
acts.  The left and right equations of motion are equations on the 
extended Keldysh contour of Fig. \ref{ksv}.c and $\d(z-z')=\frac{\dr}{\dr z}
\theta(z,z')=-\frac{\dr}{\dr z'}\theta(z,z')$.
For any $z\neq z'$, the equations of motion are solved by the 
evolution operator on the contour
$\bS(z;z')=T_{\rm K}\{{\rm e}^{-i\int_{z'}^{z}\dr\zb 
\bH_{s}(\zb)}\}$, 
since $i\frac{\overrightarrow{\dr}}{\dr z}\bS(z;z')=\bH_{s}(z)\bS(z;z')$ 
and $\bS(z;z')(-i\frac{\overleftarrow{\dr}}{\dr z'})=\bS(z;z')\bH_{s}(z')$.
Therefore, any Green function
\be
\bcalG(z;z')=\theta(z,z')\bS(z;0_{-})\bbf^{>}\bS(0_{-};z')
+\theta(z',z)\bS(z;0_{-})\bbf^{<}\bS(0_{-};z'),
\ee
with $\bbf^{\lessgtr}$ constrained by
$\bbf^{>}-\bbf^{<}=-i\bone$,
is solution of Eqs. (\ref{eomg}).
In order to fix the matrix $\bbf^{>}$ 
or $\bbf^{<}$ we impose the KMS boundary conditions. The matrix 
$\bH_{s}(z)=\bH_{s}$ for any $z$ on the vertical track, meaning 
that $\bS(-i\b;0_{-})={\rm e}^{-\b\bH_{s}}$. Eq. (\ref{bclr}) then 
implies $\bbf^{<}=-{\rm e}^{-\b(\bH_{s}-\m)}\bbf^{>}$, and taking into 
account the constraint $\bbf^{>}-\bbf^{<}=-i\bone$ we conclude that 
$\bbf^{<}=if(\bH_{s})$, 
where $f(\w)=1/[{\rm e}^{\b(\w-\m)}+1]$ is the Fermi distribution function. 
The matrix $\bbf^{>}$ takes the form $\bbf^{>}=i[f(\bH_{s})-\bone]$.

The Green function $\bcalG(z;z')$ for a system of non-interacting 
electrons is now completely fixed. Let us consider some Keldysh-Green 
functions. For $z=t_{+}$ and $z'=t_{-}$ we have the {\em greater} Green 
function while for $z=t_{-}$ and $z'=t_{+}$ we have the {\em lesser} 
Green function
\be
\bcalG^{>}(t;t')=i\,\bS(t;0)
[f(\bH_{s})-\bone]\bS(0;t'),
\quad
\bcalG^{<}(t;t')=i\,\bS(t;0)
f(\bH_{s})\bS(0;t').
\ee
Both $\bcalG^{>}$ and $\bcalG^{<}$ depend on the initial distribution 
function $f(\bH_{s})$. The 
diagonal matrix element of $-i\bcalG^{<}$ is nothing but the 
time-dependent value of the local electron density $n(\br,t)$, 
while $i\bcalG^{>}$ gives the local density of holes. Another way of 
writing $-i\bcalG^{<}$ is in terms of the eigenstates $|\q_{s}(0)\ket$ 
of $\bH_{s}$ with eigenvalues $\ve_{s}$. From the time-evolved eigenstate 
$|\q_{s}(t)\ket=\bS(t;0)|\q_{s}(0)\ket$ 
we can calculate the time-dependent wavefunction 
$\q_{s}(\br,t)=\bra\br|\q_{s}(t)\ket$. Inserting 
$\sum_{s}|\q_{s}(0)\ket\bra\q_{s}(0)|$ in the expression for $\bcalG^{<}$ we find
$-i\bra\br|\bcalG^{<}(t;t')|\br'\ket=
\sum_{s}f(\ve_{s})\q_{s}(\br,t)\q_{s}^{\ast}(\br',t')$, 
which for $t=t'$ reduces to the time-dependent density matrix. 
Knowing the greater and lesser Green functions we can also 
calculate $\bcalG^{\rm R,A}$. Taking into account the definitions 
(\ref{kra}) we find
\be
\bcalG^{\rm R}(t;t')=-i\theta(t-t')\bS(t;t'),
\quad
\bcalG^{\rm A}(t;t')=i\theta(t'-t)\bS(t;t')=
[\bcalG^{\rm R}(t';t)]^{\dag}.
\label{gagr}
\end{equation}
In the above expressions for $\bcalG^{\rm R,A}$ the Fermi distribution function has 
disappeared. The information carried by $\bcalG^{\rm R,A}$ is the same 
contained in the one-particle evolution operator. There is no 
information on how the system is prepared (how many particles, 
how they are distributed, etc). We use this observation to rewrite 
$\bcalG^{\lessgtr}$ in terms of $\bcalG^{\rm R,A}$
\begin{equation}
\bcalG^{\lessgtr}(t;t')=\bcalG^{\rm R}(t;0)\bcalG^{\lessgtr}(0;0)
\bcalG^{\rm R}(0;t').
\label{r<a}
\end{equation}
Thus, $\bcalG^{\lessgtr}$ is completely known once we know how to 
propagate the one-electron orbitals in time and how they are 
populated before the system is perturbed.\cite{cini,blandin} 
For later purposes, we also 
observe that an analogous relation holds for $\bcalG^{\rceil,\lceil}$
\begin{equation}
\bcalG^{\rceil}(t;\t)=i\bcalG^{\rm R}(t;0)\bcalG^{\rceil}(0;\t),
\quad
\bcalG^{\lceil}(\t;t)=-i\bcalG^{\lceil}(\t;0)\bcalG^{\rm A}(0;\t).
\label{rrlceil}
\end{equation}

\subsection{Total current using TDDFT}
\label{tcdft}

The fictitious $\bcalG$ of the KS system will in general not give correct 
one-particle properties. However by definition $\bcalG^{<}$ gives 
the correct density $n(\br,t)=-i\bra\br| \bcalG^{<}(t;t)|\br\ket$.
Also total currents are correctly given by TDDFT. If for instance 
$I_{\a}$ is the total current from a particular region $\a$ we have  
\begin{equation}
I_{\a}(t)=-e\int_{\a}\dr \br\;\frac{{\rm d}}{{\rm d}t}n(\br,t)=
e\int_{\a} {\rm d}\br
\;i\frac{{\rm d}}{{\rm d}t}\langle\br
|\bcalG^{<}(t;t)|\br\rangle.
\label{cudft}
\end{equation}
where the space integral extends over the region $\a$ ($e$ 
is the electron charge). We stress here that $I_{\a}$ is the electronic 
current (the direction of the current 
coincides with the direction of the electrons).

At this point, it is convenient to partition the system into three main regions: 
a central region $C$ consisting of the junction and a few atomic layers of the left 
and right electrodes and two regions $L$, $R$ which describe the left 
and right bulk electrodes. According to this partitioning, the KS 
Hamiltonian $\bH_{s}$ can be written as a $3\times 3$ block matrix, and 
the left equation of motion in (\ref{eomg}) reads
\begin{equation}
\left\{i\frac{{\rm d}}{{\rm d} z}\bone
-\left[
\begin{array}{ccc}
\bH_{LL}(z) & \bH_{LC} & 0 \\
\bH_{CL} & \bH_{CC}(z) & \bH_{CR} \\
0 & \bH_{RC} & \bH_{RR}(z)
\end{array}
\right]
\right\}\bcalG(z;z')
=\d(z-z')
\bone,
\label{tdse}
\end{equation}
with
\be
\bcalG(z;z')=
\left[
\begin{array}{ccc}
\bcalG_{LL}(z;z') & \bcalG_{LC}(z;z') & \bcalG_{LR}(z;z') \\
\bcalG_{CL}(z;z') & \bcalG_{CC}(z;z') & \bcalG_{CR}(z;z') \\
\bcalG_{RL}(z;z') & \bcalG_{RC}(z;z') & \bcalG_{RR}(z;z')
\end{array}
\right]
\ee
(a similar equation is easily obtained for the right equation of 
motion). Choosing $z$ on the forward branch of the Keldysh contour and $z'$ on 
the backward branch of the same contour, we obtain a left and right equation for the 
lesser Green function. 
These equations can be used to get rid of the time derivative 
in Eq. (\ref{cudft}). We find for $\a=L,R$
\begin{eqnarray}
I_{\alpha}(t)&=&e\int {\rm d}\br
\;\langle\br
|i\frac{{\rm d}}{{\rm d}t}\bcalG_{\alpha\alpha}^{<}(t;t)|\br\rangle
\nonumber \\ &=&e\int {\rm d}\br\,\langle\br|
\bH_{\alpha C}
\bcalG_{C\alpha}^{<}(t;t)-
\bcalG_{\alpha C}^{<}(t;t)
\bH_{C\alpha}|\br\rangle
=
2e\;{\rm Re}\left[{\rm Tr}_{C}\,
\left\{\mbox{\boldmath $Q$}_{\alpha}(t)
\right\}\right],
\nonumber \\
\label{totc}
\end{eqnarray}
where 
\beq
\mbox{\boldmath $Q$}_{\alpha}(t) &\equiv&
\bcalG^{<}_{C\alpha}(t;t)\bH_{\alpha C}
=\left[
\bcalG^{\rm R}(t,0) 
\bcalG^<(0,0)
\bcalG^{\rm A}(0,t)\right]_{C\alpha}
\bH_{\alpha C}
\nonumber \\ 
&=&
\bcalG_{CC}^{\rm R}(t;0)\bcalG_{CC}^{<}(0;0)
\bcalG_{C\a}^{\rm A}(0;t)\bH_{\a C}
\nonumber \\ &+&
\sum_{\b=L,R}\bcalG_{C\b}^{\rm R}(t;0)\bcalG_{\b C}^{<}(0;0)
\bcalG_{C\a}^{\rm A}(0;t)\bH_{\a C}
\nonumber \\ 
&+&
\sum_{\g=L,R}\bcalG_{CC}^{\rm R}(t;0)\bcalG_{C\g}^{<}(0;0)
\bcalG_{\g\a}^{\rm A}(0;t)\bH_{\a C}
\nonumber \\ &+&
\sum_{\b\g=L,R}\bcalG_{C\b}^{\rm R}(t;0)\bcalG_{\b\g}^{<}(0;0)
\bcalG_{\g\a}^{\rm A}(0;t)\bH_{\a C}
\label{qa}
\eeq
is a one-particle operator in the central region $C$ and 
${\rm Tr}_{C}$ denotes the trace over a complete set of 
one-particle states of $C$. Let us express the quantity $\bQ_{\a}$ in 
terms of the Green function $\bcalG_{CC}$ projected in the central 
region. We introduce the uncontacted Green function 
$\mbox{\boldmath $g$}$ 
which obeys Eqs. (\ref{eomg}) with $\bH_{\a C}=\bH_{C\a}=0$,
\begin{equation}
\left\{i\frac{{\rm d}}{{\rm d} z}\bone
-\left[
\begin{array}{ccc}
\bH_{LL}(z) & 0 & 0 \\
0 & \bH_{CC}(z) & 0 \\
0 & 0 & \bH_{RR}(z)
\end{array}
\right]
\right\}\mbox{\boldmath $g$}(z;z')
=\d(z-z')
\bone,
\label{ueomg}
\end{equation}
where
\be
\mbox{\boldmath $g$}(z;z')=
\left[
\begin{array}{ccc}
\mbox{\boldmath $g$}_{LL}(z;z') & 0 & 0 \\
0 & \mbox{\boldmath $g$}_{CC}(z;z') & 0 \\
0 & 0 & \mbox{\boldmath $g$}_{RR}(z;z')
\end{array}
\right]
\ee
and the same KMS boundary conditions as $\bcalG$. The uncontacted 
$\mbox{\boldmath $g$}$ allows us to convert Eqs. (\ref{eomg}) into an 
integral equation which entails the KMS boundary conditions for $\bcalG$
\begin{eqnarray}
\bcalG(z;z')&=&\mbox{\boldmath $g$}(z;z')+
\int_{\gamma}{\rm d}\bar{z} \,
\mbox{\boldmath $g$}(z;\bar{z})
\bH_{\rm off}\bcalG(\bar{z};z')
\nonumber \\ &=&
\mbox{\boldmath $g$}(z;z')+
\int_{\gamma}{\rm d}\bar{z} \,
\bcalG(z;\bar{z})
\bH_{\rm off}\mbox{\boldmath $g$}(\bar{z};z'),
\label{de}
\end{eqnarray}
$\gamma$ being the extended Keldysh contour of Fig. \ref{ksv}.c and 
$\bH_{\rm off}$ is the off-diagonal part of $\bH_{s}$.
Using the relations (\ref{cra}) of Section \ref{kbk} we find
\be
\bcalG^{\rm R,A}_{C\a}=\bcalG^{\rm R,A}_{CC}\cdot\bH_{C\a}
\bg^{\rm R,A}_{\a\a},
\quad 
\bcalG^{\rm A}_{\b\a}=\d_{\b\a}\bg^{\rm A}_{\b\b}+
\bg^{\rm A}_{\b\b}\bH_{\b C}\cdot\bcalG^{\rm A}_{CC}\cdot
\bH_{C\a}\bg^{\rm A}_{\a\a} .
\label{step1}
\ee
In Eq. (\ref{qa}) all matrix elements of $\bcalG^{<}$ are evaluated at 
times $(0;0)$. From Eq. (\ref{c><}) we see that $c^{<}(0;0)=\left[
a^{\rceil}\star b^{\lceil}\right](0;0)$, due to the theta-functions in the 
retarded and advanced components. Therefore
\be
\bcalG^{<}_{\b C}(0;0)=\left[
\bg^{\rceil}_{\b\b}\bH_{\b C}\star \bcalG_{CC}^{\lceil}\right](0;0),
\quad
\bcalG^{<}_{C\g}(0;0)=\left[
\bcalG^{\rceil}_{CC}\star\bH_{C\g}\bg_{\g\g}^{\lceil}\right](0;0),
\label{step2}
\ee
and exploiting the first two relations in Eq. (\ref{cm}) we also 
find that
\be
\bcalG^{<}_{\b\g}(0;0)=\d_{\b\g}\bg^{<}_{\b\b}(0;0)+
\left[\bg^{\rceil}_{\b\b}\bH_{\b C}\star
\bcalG_{CC}^{\rm M}\star\bH_{C\g}\bg_{\g\g}^{\lceil}\right](0;0).
\label{step3}
\ee
Substituting Eqs. (\ref{step1}-\ref{step2}-\ref{step3}) into Eq. (\ref{qa})
and using the identities (\ref{r<a}-\ref{rrlceil}) for the Green 
function $\bg$, we obtain the following expression for $\bQ_{\a}(t)$
\begin{eqnarray}
\bQ_{\alpha}(t)&=&
\sum_{\beta=L,R}
\left[
\bG^{\rm R}\cdot\bgS^{<}_{\beta}\cdot
\left(\delta_{\beta\alpha}+\bG^{\rm A}\cdot\bgS^{\rm A}_{\alpha}\right)
\right](t;t)
\nonumber \\ &+&\sum_{\beta=L,R}\left[
\bG^{\rm R}
\cdot\bgS^{\rceil}\star\bG^{\rm M}\star
\bgS^{\lceil}_{\beta}\cdot
\left(
\delta_{\beta\alpha}+
\bG^{\rm A}
\cdot\bgS^{\rm A}_{\alpha}\right)\right](t;t)
\nonumber \\
&+&
i\sum_{\beta=L,R}\bG^{\rm R}(t;0)\left[
\bG^{\rceil}
\star\bgS_{\beta}^{\lceil}
\cdot
\left(
\delta_{\beta\alpha}+
\bG^{\rm A}
\cdot\bgS_{\alpha}^{\rm A}\right)\right](0;t)
\nonumber \\
&+&
\left(
\bG^{\rm R}(t;0)\bG^{<}(0;0)
-i\left[
\bG^{\rm R}
\cdot\bgS^{\rceil}\star
\bG^{\lceil}\right](t;0)
\right)
\left[\bG^{\rm A}\cdot
\bgS_{\alpha}^{\rm A}\right](0;t),
\label{qapf}
\end{eqnarray}
where we have used the short-hand notation $\bG\equiv \bcalG_{CC}$ and 
\begin{equation}
\mbox{\boldmath $\Sigma$}(z;z')=\sum_{\alpha=L,R}
\mbox{\boldmath $\Sigma$}_{\alpha},\quad
\mbox{\boldmath $\Sigma$}_{\alpha}(z;z')=\bH_{C\alpha}\,
\mbox{\boldmath $g$}_{\a\a}(z;z')
\bH_{\alpha C}
\label{se}
\end{equation}
is the so-called embedding self-energy which accounts for hopping in 
and out of region $C$. 

Having the quantity $\bQ_{\a}(t)$ we can calculate the exact 
total current $I_{\a}(t)$ of an interacting system of electrons. 
Eq. (\ref{totc}) allows for studying transient effects and more 
generally any kind of time-dependent current responses. In the 
long time limit
\be
\lim_{t\ra\inf}\bQ_{\a}(t)=
\left[
\bG^{\rm R}\cdot\bgS^{<}_{\a}+\bG^{\rm R}\cdot\bgS^{<}
\cdot\bG^{\rm A}\cdot\bgS^{\rm A}_{\alpha}
\right](t;t)
\label{qaasym}
\ee
provided $\bG$ and $\bgS$ 
tend to zero when the separation between their time arguments increases 
(in this case, it is only the first term on the r.h.s. of Eq. 
(\ref{qapf}) that does not vanish). This condition is 
not stringent and is fulfilled provided the electrode states form a 
continuum and the local density of states in the central region $C$ is 
a smooth function. In the next Section we investigate under what 
circumstances a steady current $I_{\a}$ develops in the long-time 
limit. We will also discuss the dependence of $I_{\a}$ on the history 
of the external potential.

\subsection{Steady state and history dependence}
\label{steady}

In this Section we show that a steady state develops provided 
1) the KS Hamiltonian 
$\mbox{\boldmath $H$}_{s}(t)$  
{\em globally} converges to an asymptotic KS Hamiltonian 
$\mbox{\boldmath $H$}^{\infty}_{s}$ when $t\ra\inf$ and 
2) the electrodes form a continuum of states (thermodynamic limit) and 
the local density of states is a smooth function in the central region.

Let us define the asymptotic KS Hamiltonian of electrode $\a$ as 
$\bH_{\a\a}^{\inf}=\lim_{t\ra\inf}\bH_{\a\a}(t)$.
The retarded/advanced component of the uncontacted Green function 
$\mbox{\boldmath $g$}$ behaves like
\begin{equation}
\lim_{t\ra\inf}\mbox{\boldmath $g$}_{\a\a}^{\rm R}(t,0)=
i\,{\rm e}^{-i\bH_{\a\a}^{\inf}t}\bcalS,\quad\quad
\lim_{t\ra\inf}\mbox{\boldmath $g$}_{\a\a}^{\rm A}(0,t)=
-i\,\bcalS^{\dag}{\rm e}^{i\bH_{\a\a}^{\inf}t}
\end{equation}
where $\bcalS$ is a unitary operator and it is 
defined according to
\begin{equation}
\bcalS=
\lim_{t\ra\inf}\frac{T\left\{{\rm e}^{-i\int_{0}^{t}\bH_{\a\a}(t')dt'}\right\}}
{{\rm e}^{-i\bH_{\a\a}^{\inf}t}},
\end{equation}
$T$ being the time-ordering operator. In terms of diagonalising 
one-body states $|\psi^{\infty}_{m\alpha}\rangle$ of 
$\bH^{\infty}_{\alpha\alpha}$ with 
eigenvalues $\ve^{\infty}_{m\alpha}$, the lesser component of the 
embedding self-energy, defined in Eq. (\ref{se}), can be written as
\begin{eqnarray}
\lim_{t,t'\ra\inf}\mbox{\boldmath $\Sigma$}^{<}_{\alpha}(t;t')&=&
\lim_{t,t'\ra\inf}
\bH_{C\alpha}\,
\mbox{\boldmath $g$}_{\a\a}^{\rm R}(t;0)
\mbox{\boldmath $g$}_{\a\a}^{<}(0;0)
\mbox{\boldmath $g$}_{\a\a}^{\rm A}(0;t')
\bH_{\alpha C}
\nonumber \\ &=&
i\sum_{m,m'}
{\rm e}^{-i[\ve^{\infty}_{m\alpha}t-\ve^{\infty}_{m'\alpha}t']}
\bH_{C\alpha}|\psi^{\infty}_{m\alpha}\rangle 
\langle\psi^{\infty}_{m\alpha}|
f(\bcalS\bH_{\a\a}(0)
\bcalS^{\dag})|\psi^{\infty}_{m'\alpha}\rangle
\langle\psi^{\infty}_{m'\alpha}|
\bH_{\alpha C},
\end{eqnarray}
where we have taken into account that $\mbox{\boldmath $g$}_{\a\a}^{<}(0;0)
=if(\bH_{\a\a}(0))$.
The left and right contraction with a nonsingular hopping matrix
$\bH_{\a C}$ causes a perfect destructive interference for states with 
$|\ve^{\infty}_{m\alpha}-\ve^{\infty}_{m'\alpha}| \gtrsim 1/(t+t')$ and hence the 
restoration  
of translational invariance in time
\begin{equation}
\lim_{t,t'\ra\inf}\mbox{\boldmath $\Sigma$}^{<}_{\alpha}(t;t')=i\sum_{m}f_{m\alpha}
\mbox{\boldmath $\Gamma$}_{m\alpha}{\rm e}^{-i \ve^{\infty}_{m\alpha}(t-t')},
\label{asyse}
\end{equation}
where $f_{m\alpha}=
\langle\psi^{\infty}_{m\alpha}|f(\bcalS\bH_{\a\a}(0)
\bcalS^{\dag})
|\psi^{\infty}_{m\alpha}\rangle$ 
while 
$
\mbox{\boldmath $\Gamma$}_{m\alpha}=
\bH_{C\alpha}|\psi^{\infty}_{m\alpha}\rangle 
\langle\psi^{\infty}_{m\alpha}|
\bH_{\alpha C}$.
In principle, there may be degeneracies which require a diagonalisation
to be performed for states on the energy shell.
The above \textit{dephasing mechanism} is the key ingredient for 
a steady state to develop. Substituting Eq. (\ref{asyse}) into 
Eq. (\ref{qaasym}) we obtain for the steady state current 
\begin{eqnarray}
I_{\alpha}^{\rm (S)}=&-&2e\sum_{m\beta}f_{m\beta}
{\rm Tr}_{C}\,\left\{
\mbox{\boldmath $G$}^{\rm R}(\ve^{\infty}_{m\beta})\mbox{\boldmath $\Gamma$}_{m\beta}
\mbox{\boldmath $G$}^{\rm A}(\ve^{\infty}_{m\beta})
{\rm Im}[\mbox{\boldmath $\Sigma$}^{\rm A}_{\alpha}(\ve^{\infty}_{m\beta})]
\right\}\nonumber \\
&-&2e\sum_{m}f_{m\a}{\rm Tr}_{C}\,\left\{\mbox{\boldmath $\Gamma$}_{m\alpha}
{\rm Im}[\mbox{\boldmath $G$}^{\rm R}(\ve^{\infty}_{m\alpha})]
\right\}
\label{current4}
\end{eqnarray}
with
\be
\mbox{\boldmath $G$}^{\rm R,A}(\varepsilon)=
\frac{1}{
\varepsilon{\bf 1}_{C}-\bH^{\infty}_{CC}-
\mbox{\boldmath $\Sigma$}^{\rm 
R,A}(\varepsilon)}.
\label{asygra}
\ee
The imaginary part of $\bG^{\rm R}$ is simply given by 
$\bG^{\rm R}{\rm Im}[\bgS^{\rm R}]\bG^{\rm A}$. By definition we have
\be
\bgS^{\rm R,A}_{\a}(\ve)=
\bH_{C\a}\frac{1}{\ve{\bf 1}_{\a}-\bH_{\a\a}^{\inf}\pm i\eta}\bH_{\a C}
\ee
and hence
\be
{\rm Im}\left[\bgS^{\rm R,A}_{\a}(\ve)\right]=
\mp \p\sum_{m}\d(\ve-\ve_{m\a}^{\inf})\bgG_{m\a}.
\label{rise}
\ee
Using the above identity, the steady-state current can be rewritten in 
a Landauer-like \cite{l1957} form 
\begin{equation}
I^{(\rm S)}_{R}=-e\sum_{m}[f_{mL}\mathcal{T}_{mL}-f_{mR}\mathcal{T}_{mR}]=
-I^{(\rm S)}_{L}.
\label{current5}
\end{equation}
In the above formula $\mathcal{T}_{mR}=\sum_{n}\mathcal{T}_{mR}^{nL}$ and 
$\mathcal{T}_{mL}=\sum_{n}\mathcal{T}_{mL}^{nR}$ are the TDDFT transmission coefficients
expressed in terms of the quantities 
\begin{equation}
\mathcal{T}_{m\alpha}^{n\beta}=
2\pi\delta(\ve^{\infty}_{m\alpha}-\ve^{\infty}_{n\beta}){\rm Tr}_{C}\,\left\{
\mbox{\boldmath $G$}^{\rm R}(\ve^{\infty}_{m\alpha})
\mbox{\boldmath $\Gamma$}_{m\alpha}
\mbox{\boldmath $G$}^{\rm A}(\ve^{\infty}_{n\beta})\mbox{\boldmath $\Gamma$}_{n\beta}
\right\}=\mathcal{T}_{n\beta}^{m\alpha}.
\end{equation}

Despite the formal analogy with the Landauer formula, 
Eq. (\ref{current5}) contains an important conceptual difference since
$f_{m\alpha}$ is not simply 
given by the Fermi distribution function. For example, if the 
induced change in effective potential varies widely in space 
deep inside the electrodes, the band structure of the $\a$-electrode 
Hamiltonian $\bcalS\bH_{\a\a}(0)
\bcalS^{\dag}$ 
might differ from that of 
$\bH^{\infty}_{\alpha\alpha}$. 
However, for metallic electrodes with a macroscopic cross section
the switching on of an electric field excites 
plasmon oscillations which dynamically screen the external disturbance. 
Such a metallic screening prevents any rearrangements of the 
initial equilibrium bulk-density, provided the 
time-dependent perturbation is slowly varying during a typical 
plasmon time-scale (which is usually less than a fs). Thus,  
the KS potential $v_{s}$ undergoes a 
uniform time-dependent shift deep inside the left and right 
electrodes and the KS potential-drop is 
entirely limited to the central region. 
Denoting with $\D v_{\a}(t)$ the difference in electrode 
$\a$ between the KS potential at time $t$ and the KS potential at 
negative times, 
$\D v_{\a}(t)=v_{s}({\bf r}\in\a,t)-v_{s}({\bf r}\in\a,0)$, 
to leading order in $1/N$ we then have  
\begin{equation}
\bH_{\alpha\alpha}(t)=
\bH_{\alpha\alpha}(0)+{\bf 1}_{\a}
\D v_{\a}(t),
\label{asytddft}
\end{equation}
meaning that $\bH^{\infty}_{\alpha\alpha}=
\bH_{\alpha\alpha}(0)+{\bf 1}_{\a}\D v_{\a}^{\inf}$. 
Hence, except for corrections which are of lower order 
with respect to the system size, 
$\bcalS\bH_{\a\a}(0)\bcalS^{\dag}=
\bH_{\a\a}(0)$ and 
\begin{equation}
f_{m\alpha}=f(\ve^{\infty}_{m\alpha}-\D v_{\a}^{\inf}).
\label{ferd}
\end{equation}  
The formula for the current can be further manipulated when 
Eq. (\ref{ferd}) holds. Let us write the embedding self-energy as the 
sum of a real and imaginary part 
$\bgS_{\a}^{\rm R,A}(\ve)=\bgL_{\a}(\ve)\mp i\bgG_{\a}(\ve)/2$. 
Using Eq. (\ref{rise}) we can rewrite the transmission coefficients as
\be
\mathcal{T}_{mR}={\rm Tr}_{C}\,\left\{
\bG^{\rm R}(\ve^{\infty}_{mR})
\bgG_{mR}
\bG^{\rm A}(\ve^{\infty}_{mR})
\bgG_{L}(\ve^{\infty}_{mR})
\right\},
\ee
\be
\mathcal{T}_{mL}={\rm Tr}_{C}\,\left\{
\bG^{\rm R}(\ve^{\infty}_{mL})
\bgG_{mL}
\bG^{\rm A}(\ve^{\infty}_{mL})
\bgG_{R}(\ve^{\infty}_{mL})
\right\}.
\ee
Substituting these expressions in Eq. (\ref{current5}) and taking into 
account Eq. (\ref{ferd}) we obtain
\begin{equation}
I^{(\rm S)}_{R}=
-e\int\frac{d\ve}{2\p}\left[
f(\ve-\D v_{L}^{\inf})-f(\ve-\D v_{R}^{\inf})\right]
{\rm Tr}_{C}\,\left\{
\bG^{\rm R}(\ve)
\bgG_{L}(\ve)
\bG^{\rm A}(\ve)\bgG_{R}(\ve)
\right\}.
\label{current6}
\end{equation}

In the above equation the Green functions are calculated from 
Eq. (\ref{asygra}). The Hamiltonian $\bH_{CC}^{\inf}$ is the KS 
Hamiltonian $\bH_{s}(t\ra\inf)$ projected on region $C$ and can be 
obtained by evolving the system for very long times. 
According to the Runge-Gross theorem, $\bH_{CC}^{\inf}$ depends on 
how the system was prepared at $t=0$ (in our case the system is 
contacted and in thermal equilibrium) and on the full history of 
the time-dependent density. Therefore, {\em the use of Eq. (\ref{current6}) 
in the context of static DFT is generally not correct}. Indeed, static 
DFT is an equilibrium theory while here we are dealing with a 
non-equilibrium process. 
One might argue that in the linear-response regime the static DFT approach is 
free from the above criticism. Unfortunately, this is not the case. 
Denoting with $\d v_{\a}^{\inf}$ the 
small change $\D v_{\a}^{\inf}$ of the effective potential in 
electrode $\a$ and with $\d I^{(\rm S)}_{R}$ the corresponding 
current response, to first order in $\d v_{\a}^{\inf}$ Eq. (\ref{current6}) 
yields
\begin{equation}
\d I^{(\rm S)}_{R}=
-e\int\frac{d\ve}{2\p}\frac{\de f(\ve)}{\de \ve}
{\rm Tr}_{C}\,\left\{
\bG^{\rm R}_{0}(\ve)
\bgG_{0,L}(\ve)
\bG^{\rm A}_{0}(\ve)\bgG_{0,R}(\ve)
\right\}
\left(\d v_{R}^{\inf}- \d v_{L}^{\inf}\right).
\label{current7}
\end{equation}
The Green functions and the $\bgG$'s in Eq. (\ref{current7}) 
refer to the system in equilibrium and 
static DFT approaches can be used to evaluate the trace. 
However, DFT is not enough to calculate the change 
$\d v_{\a}^{\inf}$. Indeed
\be
\d v_{\a}^{\inf}=\lim_{t\ra\inf}\lim_{x\ra\pm\inf}
\left[\d v_{\rm ext}(\br,t)+\d V_{\rm H}(\br,t)+
\d v_{\rm xc}(\br,t)\right],
\label{lrveff}
\ee
where $x$ is the longitudinal coordinate, the plus sign applies 
for $\a=R$ and the minus sign for $\a=L$. In the above equation $v_{\rm 
ext}$ is the external potential and $V_{\rm H}$ is the Hartree 
potential; their sum gives the electrostatic Coulomb
potential $v_{\rm C}$,
\be
\d v_{\a,\rm C}=\lim_{t\ra\inf}\lim_{x\ra\pm\inf}
\left[\d v_{\rm ext}(\br,t)+\d V_{\rm H}(\br,t)\right].
\ee
The variation $\d v_{\rm xc}$ of the 
exchange-correlation potential can be expressed in terms of the exchange-correlation 
kernel $f_{\rm xc}(\br,t;\br',t')=\d v_{\rm xc}(\br,t)/\d n(\br',t')$
\be
\d v_{\a,\rm xc}=\lim_{t\ra\inf}\lim_{x\ra\pm\inf}\d v_{\rm xc}(\br,t)=
\lim_{t\ra\inf}\lim_{x\ra\pm\inf}
\int\dr \br' \int\dr t'
f_{\rm xc}(\br,t;\br',t')\d n(\br',t').
\ee
The kernel $f_{\rm xc}$ depends only on the difference $t-t'$. We 
denote by $f_{\a,\rm xc}(\br',\w)$ the Fourier transform of 
$f_{\rm xc}$ evaluated at $x=\pm\inf$ for $\a=R,L$. Then
\be
\d v_{\a,\rm xc}
=\lim_{t\ra\inf}\int \frac{d\w}{2\p}{\rm e}^{-i\w t}
\int \dr \br' f_{\a,\rm xc}(\br',\w)\d n(\br',\w)
\label{fxc}
\ee
with $\d n(\br,\w)$ the Fourier transform of $\d n(\br,t)$. 
Rewriting $\d v_{\a}^{\inf}$ as $\d v_{\a,\rm C}+\d v_{\a,\rm xc}$ and 
taking into account Eq. (\ref{fxc}), the current response $\d I^{(\rm S)}_{R}$ 
in Eq. (\ref{current7}) can also be written as
\begin{eqnarray}
\d I^{(\rm S)}_{R}=
-e\int\frac{d\ve}{2\p}
\frac{\de f(\ve)}{\de \ve}
T(\ve)
&&\left[
\left(\d v_{R,\rm C}- \d v_{L,\rm C}\right)+
\lim_{t\ra\inf}\int \frac{d\w}{2\p}{\rm e}^{-i\w t}\right.
\nonumber \\ 
\times&&\left.
\int \dr \br' 
\left(f_{R,\rm xc}(\br',\w)-f_{L,\rm xc}(\br',\w)\right)
\d n(\br',\w)
\right]
\nonumber \\
\label{current8}
\end{eqnarray}
with $T(\ve)={\rm Tr}_{C}\,\left\{
\bG^{\rm R}_{0}(\ve)
\bgG_{0,L}(\ve)
\bG^{\rm A}_{0}(\ve)\bgG_{0,R}(\ve)
\right\}$.
At zero temperature $\de f(\ve)/\de \ve=\d(\ve-\ve_{F})$, with 
$\ve_{F}$ the Fermi energy, and Eq. (\ref{current8}) becomes
\begin{eqnarray}
\d I^{(\rm S)}_{R}=
G_{\rm KS}(\ve_{F})
&&\left[
\left(\d v_{R,\rm C}- \d v_{L,\rm C}\right)+
\lim_{t\ra\inf}\int \frac{d\w}{2\p}{\rm e}^{-i\w t}\right.
\nonumber \\ 
\times&&\left.
\int \dr \br' 
\left(f_{R,\rm xc}(\br',\w)-f_{L,\rm xc}(\br',\w)\right)
\d n(\br',\w)
\right]
\label{current9}
\end{eqnarray}
where $G_{\rm KS}(\ve_{F})=-eT(\ve_{F})/2\p$ is the conductance of the 
KS system. We conclude that {\em also in the linear-response regime static DFT is 
not appropriate} for calculating the conductance since dynamical 
exchange-correlation effects might contribute through the last term 
in Eq. (\ref{current9}). Eq. (\ref{current9}) can also be obtained within 
the framework of time-dependent current density functional theory as it 
has been shown in Ref. \onlinecite{bke2005}.

We emphasize that the steady-state current in 
Eq. (\ref{current5}) results from a pure dephasing mechanism in the 
fictitious noninteracting problem. 
The damping effects of scattering are described by 
$A_{\rm xc}$ and $v_{\rm xc}$. Furthermore, the current depends 
only on the asymptotic value of the 
KS potential, $v_{s}({\bf r},t\rightarrow\infty)$.
However, $v_{s}({\bf r},t\rightarrow\infty)$ might depend on the history of the 
external applied potential and the resulting steady-state 
current might be history dependent. In these cases the full time 
evolution can not be avoided.
In the case of Time Dependent Local Density 
Approximation (TDLDA), the exchange-correlation potential $v_{\rm xc}$ 
depends only locally on the instantaneous density and has no memory at all. 
If the density tends to a constant, so does the KS potential 
$v_{s}$, which again implies that the density tends to a constant. Owing 
to the non-linearity of the problem there might still be more than one steady-state 
solution or none at all. 
We are currently investigating the possibility of  having more than one 
steady state solution.

\section{Quantum transport: A practical scheme based on TDDFT}
\label{pract}

The theory presented in the previous Sections allows us to calculate the 
time-dependent current in terms of the Green function $\bcalG_{CC}=\bG$ projected 
in the central region. In practise, it is computationally very expensive 
to propagate $\bG(z;z')$ in time (because it depends on two time 
variables) and also calculate $\bQ_{\a}$ from Eq. (\ref{qapf}). Here 
we describe a feasible numerical scheme based on the propagation of KS 
orbitals. We remind the reader that our electrode-junction-electrode 
system is infinite and non-periodic. Since one can in practice only deal 
with finite systems we will propagate KS orbitals projected in the 
central region $C$ by applying the correct boundary conditions.\cite{ksarg2005}

We specialize the discussion to nonmagnetic systems at zero temperature 
and we denote with $\q_{s}(\br,0)\equiv\bra\br|\q_{s}(0)\ket$ the eigenstates 
of $\bH_{s}(t<0)$. The time dependent density can be computed in the usual way by 
$n(\br,t)=\sum_{\rm occ}|\q_{s}(\br,t)|^{2}$, where the sum is over 
the occupied Kohn-Sham orbitals and 
$|\q_{s}(t)\ket$ is the solution of the KS equation of 
TDDFT $i\frac{\dr}{\dr t}|\q_{s}(t)\ket=\bH_{s}(t)|\q_{s}(t)\ket$. 
Using the continuity equation, we can write the total current 
$I_{\a}(t)$ of Eq. (\ref{cudft}) as 
\beq
I_{\a}(t)&=&-e\sum_{\rm occ}\int_{\a}\dr\br \grad\cdot 
{\rm Im}\left[\q_{s}^{\ast}(\br,t)\grad\q_{s}(\br,t)\right]
\nonumber \\ &=&
-e\sum_{\rm occ}\int_{S_{\a}}\dr\s\,\hat{{\bf n}}\cdot 
{\rm Im}\left[\q_{s}^{\ast}(\br,t)\grad\q_{s}(\br,t)\right]
\label{tdcoa}
\eeq
where $\hat{{\bf n}}$ is the unit vector perpendicular to the surface 
element ${\rm d}\s$ and the surface $S_{\a}$ is perpendicular to the longitudinal 
geometry of our system. From Eq. (\ref{tdcoa}) we conclude that in 
order to calculate $I_{\a}(t)$ we only need to know the time-evolved KS 
orbitals in region $C$. This is possible provided we know 
the dynamics of the remote parts of the system. As at the end of 
Section \ref{steady}, we restrict ourselves to metallic electrodes. Then, 
the external potential and the disturbance introduced by the device region 
are screened deep inside the electrodes.
As the system size increases, the remote parts are less disturbed by the 
junction and the density inside the electrodes approaches  
the equilibrium bulk-density. Thus, the macroscopic size of the 
electrodes leads to an enormous simplification since the initial-state  
self-consistency is not disturbed far away from the constriction. 
Partitioning the KS Hamiltonian as in Eq. (\ref{tdse}), the 
time-dependent Schr\"odinger equation reads
\begin{equation}
i\frac{\dr}{\dr t}\left[\begin{array}{c}
|\q_{L}\ket \\ |\q_{C}\ket \\ |\q_{R}\ket
\end{array}
\right]=
\left[
\begin{array}{ccc}
\bH_{LL} & \bH_{LC} & 0 \\
\bH_{CL} & \bH_{CC} & \bH_{CR} \\
0 & \bH_{RC} & \bH_{RR}
\end{array}
\right]
\left[\begin{array}{c}
|\q_{L}\ket \\ |\q_{C}\ket \\ |\q_{R}\ket
\end{array}
\right],
\label{tdseq}
\end{equation}
where $|\q_{\a}\ket$ is the projected wave-function onto the region $\a=L,R,C$. 
We can solve the differential equation for $\q_{L}$ and $\q_{R}$ 
in terms of the retarded Green function $\bg^{\rm R}_{\a\a}$. 
Then, we have for $\a=L,R$
\begin{equation}
|\q_{\a}(t)\ket=i\bg^{\rm R}_{\a\a}(t,0)|\q_{\a}(0)\ket+
\int_{0}^{t}{\rm d}t'\bg^{\rm R}_{\a\a}(t,t')
\bH_{\a C}|\q_{C}(t')\ket.
\label{leadeq}
\end{equation}
Using Eq. (\ref{leadeq}), the equation for $\q_{C}$ can be written as 
\beq
i\frac{\dr}{\dr t}|\q_{C}(t)\ket&=&
\bH_{CC}(t)|\q_{C}(t)\ket+\int_{0}^{t}{\rm d}t'\bgS^{\rm R}(t,t')|\q_{C}(t')\ket
 \nonumber \\ &+&i\sum_{\a=L,R}\bH_{C\a}
\bg_{\a\a}^{\rm R}(t,0)|\q_{\a}(0)\ket,
\label{evolc}
\eeq
where $\bgS^{\rm R}=\sum_{\a=L,R}\bH_{C\a}\bg^{\rm R}_{\a\a}
\bH_{\a C}$, in accordance with Eq. (\ref{se}).
Thus, for any given KS orbital we can evolve its projection 
onto the central region by solving Eq.~(\ref{evolc}) 
in region $C$. Eq. (\ref{evolc}) has also been derived elsewhere 
(for static Hamiltonians).\cite{HellumsFrensley:94} 
To summarize, all the
complexity of the infinite electrode-junction-electrode system 
has been reduced to the solution of an 
open quantum-mechanical system (the central region $C$) with proper 
time-dependent boundary conditions. 

Equation (\ref{evolc}) is the central equation of our numerical
approach to time-dependent transport. It is a reformulation of 
the original time-dependent Schr\"odinger equation (\ref{tdseq}) of the full 
system in terms of an equation for the central (device) region only. The coupling 
to the leads is taken into account by the lead Green functions 
$\bg^{\rm R}_{\a\a}, \a = L,R$. Eq. (\ref{evolc}) has the structure of a 
time-dependent Schr\"odinger equation with two extra terms. The first 
term describes the injection of particles induced by a 
non-vanishing projection of the initial wave-function onto the leads.
The second term involves the self-energy $\bgS^{\rm R}$ and the wavefunction in 
the central region at previous times during the propagation. We will 
denote it as the memory integral. We should remark here that 
these memory effects are of different origin than those which are 
usually discussed in the context of TDDFT\cite{dbg1997,vl1999}. The 
latter ones arise from the dependence of the exchange-correlation 
functional on the full history of the time-dependent density.
Most density-based functionals used at present
rely on the adiabatic approximation therefore ignoring the functional
dependence on past time-dependent densities (Ref. \onlinecite{Maitra:02}).

Equation (\ref{evolc}) is first order in time, therefore we need to specify an 
initial state which is to be propagated. We want to study the time 
evolution of systems perturbed out of their equilibrium ground state. 
Of course, the ground state of our noninteracting system is the Slater 
determinant of the occupied eigenstates of the full, extended Hamiltonian 
in equilibrium, $\bH_{s}(t<0)$. 
The practical question then arises how one can obtain these eigenstates and 
how one can propagate them in time 
without having to deal explicitly with the extended Hamiltonian. 
Below we show how we have coped with these problems.

\subsection{Computation of KS eigenstates}
\label{ks0}

Let us consider our electrode-junction-electrode system in 
equilibrium ($t<0$) and let $\q_{s}(\br)=\q_{Ej}(\br)$ be the $j$-th 
degenerate eigenstate of energy $E$ of the KS Hamiltonian $\bH_{s}$. 
The Green functions $\bcalG^{\rm R,A}(t;t')$ and $\bcalG^{<}(t;t')$ of 
the undisturbed system depend only on the difference $t-t'$. 
In absence of magnetic fields $\bH_{s}$ is invariant under 
time-reversal and the imaginary part of the Fourier transformed 
$\bcalG^{\rm R}$ is simply given by
\begin{equation}
-\frac{1}{\pi} {\rm Im} \,\left[ \bra\br|\bcalG(E)|\br'\ket \right] =  
\sum_{E'}\d(E-E')
\sum_{j=1}^{d_{E'}} \psi_{E'j}(\br) 
\psi_{E'j}^*(\br') \; .
\label{img-cont}
\end{equation}
Multiplying Eq. (\ref{img-cont}) by $\q^{\ast}_{Em}(\br)\q_{En}(\br')$ 
and integrating over $\br$ and $\br'$ in region $C$ we obtain
\be
-\frac{1}{\pi} \int_{C} \dr\br \int_{C} \dr\br'
\q^{\ast}_{Em}(\br)
{\rm Im} \,\left[ \bra\br|\bcalG(E)|\br'\ket \right]
\q_{En}(\br')
=\sum_{E'}\d(E-E')\sum_{j=1}^{d_{E'}}S_{mj}(E')S_{jn}(E'),
\label{imgrme}
\ee
where 
\be
S_{mj}(E)\equiv \int_{C} \dr\br\; \q^{\ast}_{Em}(\br)
\psi_{Ej}(\br)=S^{\ast}_{jm}(E)
\ee
is the overlap matrix in region $C$ between degenerate states. This 
matrix is Hermitian and can be diagonalized, i.e.,
\be
\sum_{j=1}^{d_{E}}S_{mj}(E)a^{(l)}_{j}(E)=\l_{l}(E)a^{(l)}_{m}(E).
\label{diag}
\ee
Next, we multiply Eq. (\ref{imgrme}) by $a^{(l)^{\ast}}_{m}(E)a^{(l')}_{n}(E)$ 
and sum over $m$ and $n$. The result can be written in terms of the new 
eigenfunctions $a_{El}(\br)=\sum_{n=1}^{d_{E}}a^{(l)}_{n}(E)\q_{En}(\br)$ as
\be
-\frac{1}{\pi} \int_{C} \dr\br \int_{C} \dr\br'
a^{\ast}_{El}(\br)
{\rm Im} \,\left[ \bra\br|\bcalG(E)|\br'\ket \right]
a_{El'}(\br')
=\d_{ll'}\l_{l}^{2}(E)
\sum_{E'}\d(E-E'),
\label{imgrme2}
\ee
where we have used Eq. (\ref{diag}) and the orthonormality of the 
$S$-matrix eigenvectors: $\sum_{j=1}^{d_{E}}a^{(l)^{\ast}}_{j}(E)a^{(l')}_{j}(E)
=\d_{ll'}$. Equation (\ref{imgrme2}) shows explicitly that the functions $a_{Ej}({\bf r})$
diagonalize ${\rm Im} \,[\bcalG_{CC}(E)]$ 
in the central region and that the eigenvalues are positive. Since 
any linear combination of degenerate eigenstates is again an eigenstate, 
diagonalizing ${\rm Im}\,[\bcalG_{CC}(E)]$ gives us one set of linearly independent, 
degenerate eigenstates of energy $E$. 
In our practical implementation described in more detail in Section 
\ref{details}, we diagonalize
\begin{equation}
-\frac{1}{\pi D_{C}(E)} {\rm Im}\,[\bcalG_{CC}(E)]
\label{img-pract}
\end{equation}
where $D_C(E) = -\frac{1}{\pi} {\rm Tr} \, 
\left\{{\rm Im}\,[\bcalG_{CC}(E)]\right\}$
is the total density of states in the central region. If we use $N_g$ grid 
points to describe the central region, the diagonalization in principle 
gives $N_g$ eigenvectors but only a few have the physical meaning of 
extended eigenstates at this energy. It is, however, very easy to identify 
the physical states by looking at the eigenvalues: at a given energy 
$E$ only $d_{E}$ eigenvalues are nonvanishing  and they always add up to unity. 
The corresponding states are the physical 
ones. All the other eigenvalues are zero (or numerically close to zero) and 
the corresponding states have no physical meaning. 

The procedure described above gives the correct extended eigenstates only up 
to a normalization factor. When diagonalizing Eq. (\ref{img-pract}) 
with typical library routines one obtains eigenvectors which are normalized 
to the central region. Physically this might be incorrect. It is 
possible to fix the normalization by  matching the wavefunction for the central 
region to the known form (and normalization) of the wavefunction in the macroscopic leads. 

It should be emphasized that the procedure described here for the extraction 
of eigenstates of the extended system from $\bcalG_{CC}(E)$ only works in practice 
if $E$ is in the continuous part of the spectrum due to the sharp peak 
of the delta function in the discrete part of the spectrum. 
Eigenstates in the discrete part of the spectrum can be found by 
considering the original Schr\"odinger equation for the full system:
$\bH_{s} \psi = E \psi$. 
Using again the block structure of the Hamiltonian this can be transformed 
into an effective Schr\"odinger equation for an {\em energy-dependent} Hamiltonian 
for the central region only:
\begin{equation}
\left( \bH_{CC} + \sum_{\a=L,R}\bH_{C\a}\frac{1}{E{\bf 1}_{\a}-\bH_{\a\a}}\bH_{\a C} \right) 
|\psi_C\ket = E |\psi_C\ket.
\label{bound}
\end{equation}
This equation has solutions only for certain values of $E$ which are 
the discrete eigenenergies of the full Hamiltonian $\bH_{s}$. Since 
the left and right electrodes form a continuum, the 
dimension of the kernel of $(E-\bH_{\a\a})$ is zero for those 
energies $E$ in the discrete part of the spectrum. We also notice that 
the second term in parenthesis in Eq. (\ref{bound}) is nothing but the real 
part of the retarded/advanced self-energy in equilibrium, see Eq. (\ref{rise}).
Bound states as well as fully reflected waves will contribute to the 
density but not to the current and might play a role
in the description of charge-accumulation in molecular transport,
as, e.g., in Coulomb blockade phenomena. In our TDDFT formulation bound 
states and fully reflected waves also play an extra role, since they are  
needed for calculating the effective potential $v_{s}$ (which is a 
functional of the density) which is in turn used for extracting all extended 
states. 

\subsection{Algorithm for the time evolution}
\label{kst}

In order to calculate the longitudinal current in an 
electrode-junction-electrode system we need to propagate the 
Kohn-Sham orbitals. The main difficulty stems from the macroscopic 
size of the electrodes whose remote parts, ultimately, are taken 
infinitely far away from the central, explicitly treated, scattering 
region $C$.

The problem can be solved by imposing transparent boundary 
conditions\cite{Moyer:04} at the electrode-junction interfaces. 
Efficient algorithms have been  
proposed for wave-packets initially {\em localized} in the scattering 
region and for Hamiltonians constant in time. In this Section 
we describe an algorithm well suited for  
delocalized initial states, as well as for localized ones, evolving 
with a time-dependent Hamiltonian. 

Let $\bH_{s}(t)$ be the time-dependent KS Hamiltonian. 
We partition $\bH_{s}(t)$ as in Section \ref{tcdft}. The explicitly treated region $C$ includes  
the first few atomic layers of the left and right electrodes. The boundaries of 
this region are chosen in such a way that the density outside $C$ 
is accurately described by an equilibrium bulk density. 
It is convenient to write $\bH_{\a\a}(t)$, with $\a=L,R$, as the sum of a 
term $\bH^{0}_{\a\a}=\bH_{\a\a}(0)$ which is constant in time and another term $\bU_{\a}(t)$ 
which is explicitly time-dependent, $\bH_{\a\a}(t)=\bH^{0}_{\a\a}+\bU_{\a}(t)$. 
In configuration space $\bU_{\a}(t)$ 
is diagonal at any time $t$ since the KS potential is local in space.
Furthermore, the diagonal elements $U_{\a}(\br,t)$ are spatially 
constant for metallic electrodes. Thus, 
$\bU_{\a}(t)=U_{\a}(t){\bf 1}_{\a}$ and $U_{L}(t)-U_{R}(t)$ is the 
total potential drop across the central region. 
We write $\bH_{s}(t)=\tilde{\bH}(t)+\bU(t)$ with 
\begin{equation}
\tilde{\bH}(t)=\left[
\begin{array}{ccc}
\bH^{0}_{LL} & \bH_{LC} & 0 \\
\bH_{CL} & \bH_{CC}(t) & \bH_{CR} \\
0 & \bH_{RC} & \bH^{0}_{RR}
\end{array}
\right],
\quad
{\rm and}
\quad
\bU(t)=\left[
\begin{array}{ccc}
U_{L}(t){\bf 1}_{L} & 0 & 0 \\
0 & 0 & 0 \\
0 & 0 & U_{R}(t){\bf 1}_{R}
\end{array}
\right].
\end{equation}
In this way, the only term  
in $\tilde{\bH}(t)$ that depends on $t$ is $\bH_{CC}(t)$. 
For any given initial state $|\q(0)\ket=|\q^{(0)}\ket$ we calculate 
$|\q(t_{m}=m\D t)\ket=|\q^{(m)}\ket$ by using a generalized form of 
the Cayley method 
\begin{equation}
\left({\bf 1}+i\d \tilde{\bH}^{(m)}\right)
\frac{{\bf 1}+i\frac{\d}{2}\bU^{(m)}}
{{\bf 1}-i\frac{\d}{2}\bU^{(m)}}|\q^{(m+1)}\ket=
\left({\bf 1}-i\d \tilde{\bH}^{(m)}\right)
\frac{{\bf 1}-i\frac{\d}{2}\bU^{(m)}}
{{\bf 1}+i\frac{\d}{2}\bU^{(m)}}|\q^{(m)}\ket,
\label{prop}
\end{equation}
with 
$\tilde{\bH}^{(m)}=\frac{1}{2}[\tilde{\bH}(t_{m+1})+\tilde{\bH}(t_{m})]$, 
$\bU^{(m)}=\frac{1}{2}[\bU(t_{m+1})+\bU(t_{m})]$ and $\d=\D t/2$. 
It should be noted that our propagator is 
norm conserving (unitary) and 
accurate to second-order in $\d$, as is the Cayley propagator.\cite{castro} 
Denoting by $|\q_{\a}\ket$ the projected wave function 
onto the region $\a=R,L,C$, we find from Eq.~(\ref{prop}) 
\begin{equation}
|\q_{C}^{(m+1)}\ket=
\frac{{\bf 1}_{C}-i\d \bH_{\rm eff}^{(m)}}{{\bf 1}_{C}+i\d \bH_{\rm eff}^{(m)}}
|\q_{C}^{(m)}\ket
+|S^{(m)}\ket-|M^{(m)}\ket.
\end{equation}
Here, $\bH_{\rm eff}^{(m)}$ is the effective Hamiltonian of the central region:
\be
\bH_{\rm eff}^{(m)}=\bH_{CC}^{(m)}-
i\d\bH_{CL}\frac{1}{{\bf 1}_{L}+i\d\bH^{0}_{LL}}\bH_{LC}
-i\d\bH_{CR}\frac{1}{{\bf 1}_{R}+i\d\bH^{0}_{RR}}\bH_{RC}
\ee
with $\bH_{CC}^{(m)}=\frac{1}{2}[\bH_{CC}(t_{m+1})+\bH_{CC}(t_{m})]$. The source term 
$|S^{(m)}\ket$ describes the injection of density into the region $C$, 
while the memory term $|M^{(m)}\ket$ is responsible for the hopping in and 
out of the region $C$. In terms of the propagator for the uncontacted 
and undisturbed $\a$ electrode
\begin{equation}
\bg_{\a}=\frac{{\bf 1}_{\a}-i\d\bH^{0}_{\a\a}}{{\bf 1}_{\a}+i\d\bH^{0}_{\a\a}},
\label{discgreen}
\end{equation}
the source term can be written as 
\begin{equation}
|S^{(m)}\ket=-\frac{2i\d}{{\bf 1}_{C}+i\d \bH_{\rm eff}^{(m)}} 
\sum_{\a=L,R}\frac{\L_{\a}^{(m,0)}}{u_{\a}^{(m)}}\bH_{C\a}
\frac{[\bg_{\a}]^{m}}{{\bf 1}_{\a}+i\d\bH^{s}_{\a\a}}|\q_{\a}^{(0)}\ket,
\label{exsrc}
\end{equation}
with
\begin{equation}
u_{\a}^{(m)}=\frac{1-i\frac{\d}{2}U_{\a}^{(m)}}{1+i\frac{\d}{2}U_{\a}^{(m)}}
\quad {\rm and}\quad
\L_{\a}^{(m,k)}=\prod_{j=k}^{m}[u_{\a}^{(j)}]^{2}.
\end{equation}
For a wave packet initially localized in $C$ the projection onto the left 
and right electrode $|\q_{\a}^{(0)}\ket$ vanishes and $|S^{(m)}\ket=0$ for any 
$m$, as it should be. The memory term is more 
complicated and reads
\begin{eqnarray}
|M^{(m)}\ket=-\frac{\d^{2}}{{\bf 1}_{C}+i\d \bH_{\rm eff}^{(m)}}\sum_{\a=L,R}\sum_{k=0}^{m-1}
\frac{\L_{\a}^{(m,k)}}{u_{\a}^{(m)}u_{\a}^{(k)}}
[\bQ_{\a}^{(m-k)}+\bQ_{\a}^{(m-k-1)}]
\nonumber \\ \times
\left(|\q_{C}^{(k+1)}\ket+|\q_{C}^{(k)}\ket\right)
\end{eqnarray}
where
\begin{equation}
\bQ_{\a}^{(m)}=\bH_{C\a}\frac{[\bg_{\a}]^{m}}{{\bf 1}_{\a}+i\d\bH^{s}_{\a\a}}\bH_{\a C}.
\label{Qdef}
\end{equation}
The quantities $\bQ_{\a}^{(m)}$ depend on the geometry of the 
system and are independent of the initial state $\q^{(0)}$. 

\begin{figure}[htbp]
\begin{center}
\includegraphics[scale=0.7]{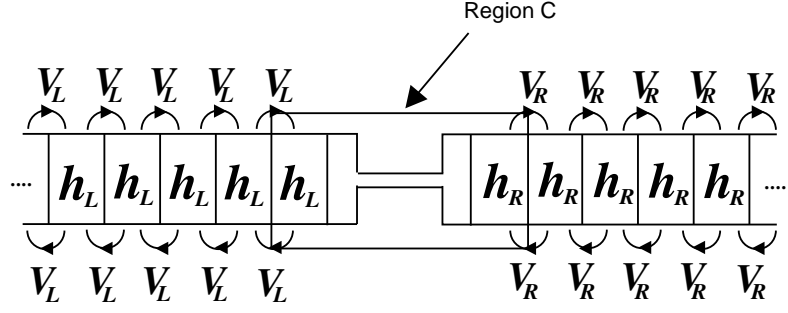}
\caption{\label{semi} Schematic sketch of an electrode-junction-electrode 
system with semiperiodic electrodes.}
\end{center}
\end{figure}

Below we propose a recursive scheme to calculate the $\bQ_{\a}^{(m)}$'s for 
those system geometries having semiperiodic electrodes along the 
longitudinal direction, see Fig. \ref{semi}. In this case $\bH^{0}_{\a\a}$ 
has a tridiagonal block form
\begin{equation}
\bH^{0}_{\a\a}=\left[
\begin{array}{cccc}
    \bh_{\a} & \bV_{\a} & 0 & \ldots \\
    \bV_{\a} & \bh_{\a} & \bV_{\a} & \ldots \\ 
    0 & \bV_{\a} & \bh_{\a} & \ldots \\ 
    \ldots & \ldots & \ldots & \ldots
\end{array}    
\right],
\end{equation}
where $\bh_{\a}$ describes a convenient cell and $\bV_{\a}$ is the 
hopping Hamiltonian between two nearest neighbor cells. 
Without loss of generality we assume that both $\bh_{\a}$ and $\bV_{\a}$ 
are square matrices of dimension $N_{\a}\times N_{\a}$. 
Taking into account that the central region 
contains the first few cells of the left and right electrodes, the 
matrix $\bQ_{\a}^{(m)}$ has the following structure
\begin{equation}
\bQ_{L}^{(m)}=\left[
\begin{array}{ccc}
\bq_{L}^{(m)} & 0 & 0 \\
0 & 0  & 0 \\
0 & 0 & 0
\end{array}    
\right],\quad\quad
\bQ_{R}^{(m)}=\left[
\begin{array}{ccc}
0 & 0 & 0 \\
0 & 0  & 0 \\
0 & 0 & \bq_{R}^{(m)}
\end{array}    
\right].
\end{equation}
The $\bq_{\a}^{(m)}$'s are square matrices of dimension $N_{\a}\times 
N_{\a}$ and are given by
\begin{equation}
\bq_{\a}^{(m)}=\bV_{\a}\left[
\frac{[\bg_{\a}]^{m}}{{\bf 1}_{\a}+i\d\bH_{\a\a}}
\right]_{1,1}\bV_{\a},
\end{equation}
where the subscript $(1,1)$ denotes the first diagonal block of the 
matrix in the square brackets. We introduce the generating matrix function
\begin{equation}
\bq_{\a}(x,y)\equiv\bV_{\a}\left[
\frac{1}{x{\bf 1}_{\a}+iy\d\bH_{\a\a}}
\right]_{1,1}\bV_{\a},
\label{q1def}
\end{equation}
which can also be expressed in terms of continued matrix fractions 
\begin{eqnarray}
\bq_{\a}(x,y)  &=& 
\bV_{\a}
\mbox{$
        \frac{
	       \mbox{ {\normalsize $1$ } } 
	      }
	      {
	       \mbox{ {\normalsize $  x+iy\d\bh_{\a}+y^{2}\d^{2}\bV_{\a}       
	                           \mbox{$
                                           \frac{
					         \mbox{ {\normalsize $1$ } } 
					        }
						{
						 \mbox{ {\normalsize $  x+iy\d\bh_{\a}+y^{2}\d^{2}\bV_{\a}
						                     \mbox{$
								             \frac{\mbox{ {\normalsize $1$ } } }{\ldots\ldots}
									   $}
						                     $ } 
						         {\normalsize $\bV_{\a}$ }
				                       }
						}
			                 $}
	                           $ } 
		        {\normalsize $\bV_{\a}$ }
	             }
	      }
      $}
\bV_{\a} 
\nonumber \\ 
&=&
\bV_{\a}\frac{1}{x+iy\d\bh_{\a}+y^{2}\d^{2}\bq_{\a}(x,y)}\bV_{\a}.
\label{qalpha}
\end{eqnarray}
The $\bq_{\a}^{(m)}$'s can be obtained from
\begin{equation}
\bq_{\a}^{(m)}=\frac{1}{m!}\left.\left[-\frac{\de}{\de x}+\frac{\de }{\de y}\right]^{m}
\bq_{\a}(x,y)\right|_{x=y=1}.
\label{gen}
\end{equation}

From Eqs.~(\ref{gen}) and (\ref{qalpha}) one can build up a recursive 
scheme. Let us define 
$\bp_{\a}^{-1}(x,y)=x+iy\d\bh_{\a}+y^{2}\d^{2}\bq_{\a}(x,y)$ and 
$\bp_{\a}^{(m)}=\frac{1}{m!}[-\frac{\de}{\de x}+\frac{\de}{\de y}]
\bp_{\a}(x,y)|_{x=y=1}$. Then, by definition,
$\bq_{\a}^{(m)}=\bV_{\a}\bp_{\a}^{(m)}\bV_{\a}$.
Using the identity $\frac{1}{m!}[-\frac{\de}{\de x}+\frac{\de}{\de y}]^{m}
\bp_{\a}(x,y)\bp_{\a}^{-1}(x,y)=0$, one finds
\begin{equation}
(1+i\d\bh_{\a})\bp_{\a}^{(m)}=(1-i\d\bh_{\a})\bp_{\a}^{(m-1)} -
\d^{2}\sum_{k=0}^{m}(\bq_{\a}^{(k)}+2\bq_{\a}^{(k-1)}+\bq_{\a}^{(k-2)})
\bp_{\a}^{(m-k)}
\label{recur}
\end{equation}
with $\bp_{\a}^{(m)}=\bq_{\a}^{(m)}=0$ for $m<0$. Once 
$\bq_{\a}^{(0)}$ has been obtained by solving Eq.~(\ref{qalpha}) 
with $x=y=1$, we can calculate $\bp_{\a}^{(0)}=
[1+i\d\bh_{\a}+\d^{2}\bq_{\a}^{(0)}]^{-1}$. Afterwards, we can use 
Eq.~(\ref{recur}) with $\bq_{\a}^{(1)}=\bV_{\a}\bp_{\a}^{(1)}\bV_{\a}$ 
to calculate $\bp_{\a}^{(1)}$ and hence $\bq_{\a}^{(1)}$ and so on and 
so forth. 

This concludes the description of our algorithm for the propagation of the 
time-dependent Schr\"odinger equation for extended systems. It is worth  
mentioning an additional complication here which arises for the propagation 
of a time-dependent Kohn-Sham equation. This complication stems from the fact 
that in order to compute $|\q_{C}^{(m+1)}\ket$ at time step $m+1$ one needs to 
know the time-dependent KS potential at the same time step which, via the 
Hartree and exchange-correlation potentials, depends on the yet unknown 
orbitals $|\q_{C}^{(m+1)}\ket$. Of course, the solution is to use a 
predictor-corrector approach: in the first step one approximates 
$\bH_{CC}^{(m)}$ by $\bH_{CC}(t_m)$, computes new orbitals 
$|\tilde{\q}_{C}^{(m+1)}\ket$ and from those obtains an improved approximation 
for $\bH_{CC}^{(m)}$.

\section{Implementation details for 1d systems and numerical results}
\label{details}

All the methodological discussion of Section \ref{pract} is general 
and can be applied to all systems having a longitudinal geometry
like the one in Fig.~\ref{semi}. In this Section we show that the 
proposed scheme is feasible by testing it against one-dimensional model 
systems.  The extension to real molecular-device configurations is presently 
under development.\cite{wirtz} We consider systems described by the 
Hamiltonian
\be
\bra x|\bH|x'\ket=\d(x-x')
\left[-\frac{1}{2}\frac{\dr}{\dr x^{2}}+V(x)
\right].
\ee
We have used a simple three-point discretization for 
the second derivative 
\begin{equation}
\frac{{\rm  d}^2}{{\rm d} x^2} \psi(x) \vert_{x=x_i} \approx 
\frac{1}{(\Delta x)^2} \left[ \psi(x_{i+1}) - 2 \psi(x_i) + 
\psi(x_{i-1}) \right]
\end{equation}
with equidistant grid points $x_i$, $i=1,\dots,N_g$ and spacing $\Delta x$. 
Within this approximation matrices of the form $\bH_{C\a} \bM \bH_{\a C}$ 
which are $N_g \times N_g$ matrices and appear, e.g., in Eq.~(\ref{se}) 
or (\ref{Qdef}), have only one nonvanishing matrix element. For $\a=L$ this 
is the $(1,1)$ element, for $\a=R$ it is the $(N_g,N_g)$ element. 

In order to proceed we have to specify the nature of the leads and therefore
the lead Green function. Here we choose the simplest case of 
semi-infinite, uniform leads at constant potential $U_{\a 0}$. In this case, 
the retarded Green function $\bg^{\rm R}_{\a\a}$ in the energy domain can be given in 
closed form:
\begin{eqnarray}
[\bg^{\rm R}_{\a\a}(E)]_{kl} &=& - \frac{i \Delta x} 
{ \sqrt{ 2 \tilde{E}_{\a} }} 
\exp\left\{i \sqrt{2 \tilde{E}_{\a} } | x_k - x_l| \right\} 
\nonumber \\ &+& \frac{i \Delta x}{\sqrt{2 \tilde{E}_{\a} }} 
\exp\left\{i \sqrt{2 \tilde{E}_{\a}}( | x_k - x_{\a0}| + |x_l - 
x_{\a0}|)\right\}
\end{eqnarray}
with $\tilde{E}_{\a} = E-U_{\a 0}$. The abscissa $x_{\a 0}$ 
is the position of the interface between the lead and the device region;
in our implementation $x_{L0}$ is the first grid point of region $C$ 
while $x_{R0}$ is the $N_{g}$-th grid point of region $C$. According 
to the notation in Eq. (\ref{tdseq}) the one-particle state of 
region $C$ describing an electron localized in $x_{L0}$ is denoted 
by $|x_{C1}\ket$ while the one-particle state of 
region $C$ describing an electron localized in $x_{R0}$ is denoted 
by $|x_{CN_{g}}\ket$. The coordinate $x_k = x_{\a 0} 
\pm k \Delta x$, $k>0$, where the plus sign applies for $\a=R$ and the minus sign 
for $\a=L$.  
\begin{figure}[htbp]
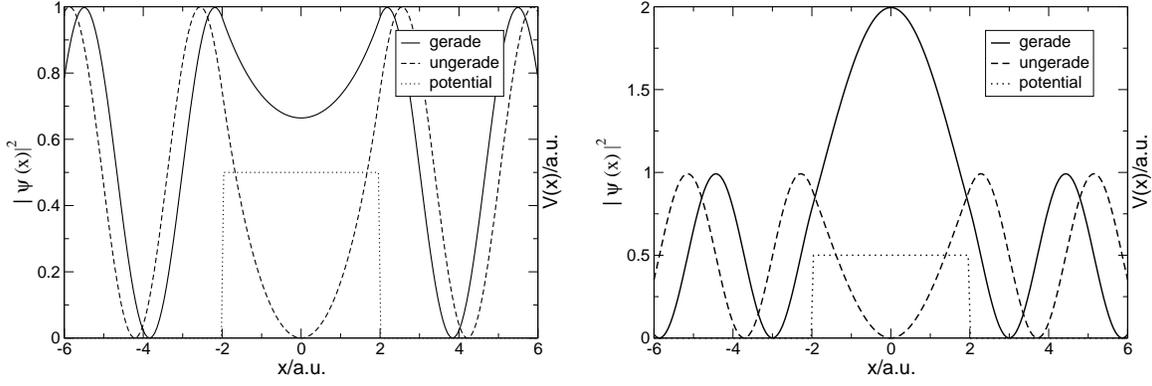

\begin{center}
\includegraphics*[scale=0.30]{fig3up}$\quad$
\includegraphics*[scale=0.30]{fig3low}\\[0.5cm]
\caption{ Continuum states of square potential barrier 
at different energies with leads at zero potential. Left panel: 
eigenstates for $\ve=0.45$ a.u., just below the barrier height of $0.5$ a.u.. 
Right panel: eigenstates at $\ve=0.6$ a.u..}
\label{ext-states}
\end{center}
\end{figure}

The results of the procedure for calculating extended eigenstates as described 
in Section \ref{ks0} is illustrated in Fig.~\ref{ext-states} for a 
square potential barrier with zero potential in both leads. 
In the left panel we 
have the square modulus of eigenstates at an energy below the barrier height 
while in the right panel eigenstates with energy higher than the barrier 
are shown. The states result from diagonalization of Eq.~(\ref{img-pract}). 
In order to obtain the normalization constant we compute the logarithmic 
derivative at the boundary of the central region numerically and match it 
to the analytic form in the lead to obtain the phase shift $\delta_{\a}$:
\begin{equation}
\frac{1}{2} \frac{{\rm d}^2}{{\rm d} x^2} \ln( |\psi(x)|^2 ) 
\bigg\vert_{x=x_{\a 0}} = q \cot(\delta_{\a})
\end{equation}
where $q= \sqrt{2 \tilde{E}_{\a}}$. Knowing the phase shift we can rescale the 
wavefunction such that it matches with the analytic form 
$\sin(q (x-x_{\a 0}) + \delta_{\a})$ at the interface. Of course, this form 
of the extended states only applies for $\tilde{E}_{\a} >0$ but as long as 
$E$ is in the continuous part of the spectrum, it is correct at least for 
one of the leads. This is sufficient to determine the normalization. The 
states obtained numerically with this procedure coincide with the known 
analytical results. 

We then implemented the propagation scheme presented in 
the previous Section. Within  our three-point approximation, 
$\bh_{\a}$, $\bV_{\a}$ and $\bq_{\a}$ are 
$1\times 1$ matrices. The equation for $q_{\a}^{(0)}$ [see 
Eqs.~(\ref{qalpha}) and (\ref{gen})] becomes a simple quadratic equation 
which can be solved explicitly
\begin{equation}
q_{\a}^{(0)} = \frac{- (1 + i \delta h_{\a}) 
+ \sqrt{(1 + i \delta h_{\a}) ^2 + 4 (\delta V_{\a})^2}}
{2 \delta^2}  \, .
\label{q0}
\end{equation}
Although the quadratic equation has two solutions, the above choice for 
$q_{\a}^{(0)}$ is dictated by the fact that the Taylor expansions for small 
$\delta$ of Eqs.~(\ref{q0}) and (\ref{qalpha}) have to coincide. 
Using this result we then 
solved the iterative scheme to obtain the $q_{\a}^{(m)}$ for $m \geq 1$. 

\begin{figure}[htbp]
\begin{center}    
\includegraphics*[scale=0.5]{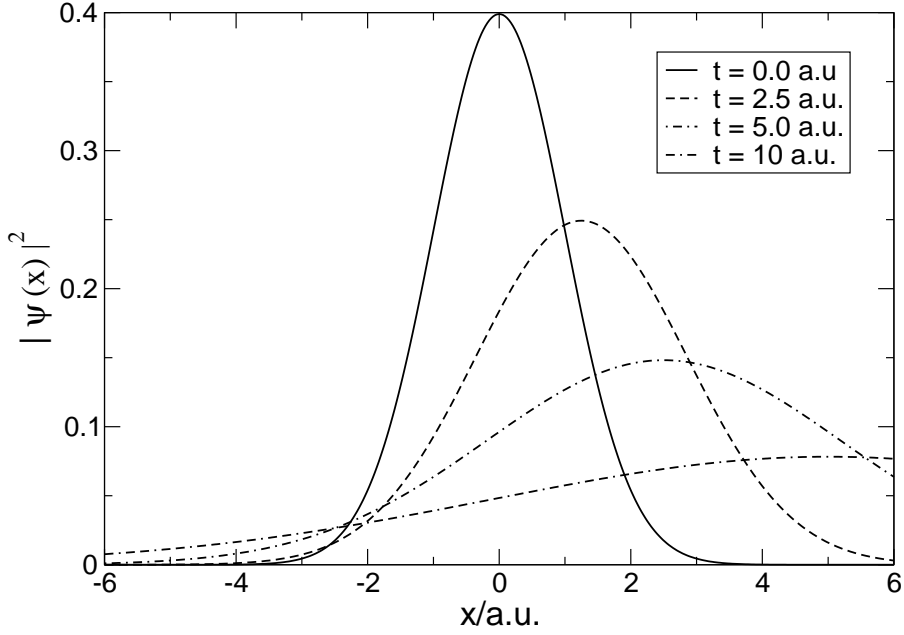}
\caption{Time evolution of a Gaussian wavepacket with 
initial width 1.0 a.u. and initial momentum 0.5 a.u. for various 
propagation times. The transparent boundary conditions allow the wavepacket 
to pass the propagation region without spurious reflections at the boundaries.}
\label{wavepacket} 
\end{center}
\end{figure}

As a first check on the propagation method we propagated a Gaussian 
wavepacket which, at initial time $t=0$, is completely localized in the 
central device region. (The source terms $|S^{(m)}\ket$ then vanish identically). 
As can be seen in Fig.~\ref{wavepacket}, the wavepacket correctly propagates 
through the boundaries without any spurious reflections. 

For the propagation of the extended initial states (the eigenstates of the 
unperturbed system) we also need to implement the source terms 
$|S^{(m)}\ket$. In 
the following we assume that the left and right leads are at the same 
potential initially so that the analytic form of the initial states 
is in both leads given by $\sin(q(x-x_{\a 0}) + \delta_{\a}) = 
\left[\exp(i\d_{\a}-iqx_{\a 0})\exp(iqx)-{\rm c.c.}\right]/2i$. 
Let us specialize the discussion to the case $\a=R$ and 
define the state $|q_{R}\ket$ according to $\bra x_{Rk}|q_{R}\ket=
\exp(iqk\D x)$, where $|x_{Rk}\ket$ is the one-particle state of 
electrode $R$ describing an electron localized in 
$x_{k}=x_{R0}+k\D x$, $k>0$. Then, the projection of the 
initial state onto lead $R$ reads $|\q_{R}^{(0)}\ket=\frac{1}{2i}
\left[\exp(i\d_{\a})|q_{R}\ket-
\exp(-i\d_{\a})|-q_{R}\ket\right]$. From Eq. (\ref{exsrc}) 
the contribution to the source term for $\a=R$ is completely known once we 
know how $\bH_{CR}[\bg_{R}]^{m}/
({\bf 1}_{R}+i\d\bH_{RR})$ acts on the state $|q_{R}\ket$. We have
\be
\bH_{CR}\frac{[\bg_{R}]^{m}}{({\bf 1}_{R}+i\d\bH_{RR})}
|q_{R}\ket=V_{R}|x_{CN_{g}}\ket\bra x_{R1}|\frac{[\bg_{R}]^{m}}{({\bf 1}_{R}+i\d\bH_{RR})}
|q_{R}\ket
\ee
where $x_{CN_{g}}$ corresponds the $N_{g}$-th discretization point of region $C$ 
(the last point on the right before electrode $R$ starts).
We rewrite the unknown quantity as follows
\be
\bra x_{R1}|
\frac{[\bg_{R}]^{m}}{{\bf 1}_{R}+i\d\bH_{RR}}|q_{R}\ket=\left.
\frac{\left[D(x,y)\right]^{m}}{m!}\r(x,y)\right|_{x=y=1},
\ee
with
\be
D(x,y)=\left(-\frac{\de}{\de x}+\frac{\de}{\de y}\right),
\quad
\r(x,y)=\bra x_{R1}|
\frac{1}{x{\bf 1}_{R}+iy\d\bcalE_{RR}}|q_{R}\ket.
\label{runk}
\ee
Next, we use the Dyson equation to find an explicit expression for 
$\r(x,y)$. We have
\be
\frac{1}{x{\bf 1}_{R}+iy\d\bH_{RR}}|q_{R}\ket=
\frac{1}{x}|q_{R}\ket
-\frac{1}{x}\frac{iy\d}{x{\bf 1}_{R}+iy\d\bH_{RR}}\bH_{RR}|q_{R}\ket.
\label{dyson}
\ee
It is straightforward to realize that the action of $\bH_{RR}$ on  
$|q_{R}\ket$ yields
\be
\bH_{RR}|q_{R}\ket=2V_{R}\cos(q\D x)|q_{R}\ket-V_{R}{\rm e}^{-iq\D 
x}|x_{R1}\ket,
\label{rrc}
\ee
so that Eq. (\ref{dyson}) can be rewritten as
\be
\left[
1+\frac{2iy\d V_{R}\cos(q\D x)}{x}
\right]
\frac{1}{x{\bf 1}_{R}+iy\d\bH_{RR}}|q_{R}\ket= \frac{1}{x}|q_{R}\ket
+\frac{1}{x}\frac{iy\d V_{R}{\rm e}^{-iq\D x}}{x{\bf 
1}_{R}+iy\d\bH_{RR}}|x_{R1}\ket.
\label{drea}
\ee
Projecting Eq. (\ref{drea}) on $\bra x_{R0}|$ we find
\be
\left[
1+\frac{2iy\d V_{R}\cos(q\D x)}{x}
\right]\r(x,y)=\frac{1}{x}+\frac{iy\d {\rm e}^{-iq\D x} }{xV_{R}}q_{R}(x,y),
\label{tbs}
\ee
where $q_{R}(x,y)$ is the generating function defined in Eq. (\ref{q1def}).
Solving Eq. (\ref{tbs}) for $\r(x,y)$ we conclude 
that
\be
V_{R}\r(x,y)=\frac{V_{R}+iy\d {\rm e}^{-iq\D x} q_{R}(x,y)}{x+2iy\d 
V_{R}\cos(q\D x)}.
\ee
Using the relation in Eq. (\ref{gen}) for the coefficients $q_{\a}^{(m)}$ 
we find
\beq
\left.
\frac{\left[D(x,y)\right]^{m}}{m!}\r(x,y)\right|_{x=y=1}&=&
\frac{\left(1-2i\d V_{R}\cos(q\D x)\right)^{m}}{\left(1+2i\d 
V_{R}\cos(q\D x)\right)^{m+1}}
+\frac{i\d}{V_{R}} {\rm e}^{-iq\D x}
\nonumber \\ &\times&\sum_{j=0}^{m}
\frac{\left(1-2i\d V_{R}\cos(q\D x)\right)^{m-j}}{\left(1+2i\d 
V_{R}\cos(q\D x)\right)^{m+1-j}}\left(q_{R}^{(j)}+q_{R}^{(j-1)}\right).
\nonumber \\
\label{totm}
\eeq
One may proceed along the same lines for extracting the left 
component of the source term.

To test our implementation we have propagated eigenstates of the extended 
system. As expected, these states just pick up an exponential phase factor 
$\exp(-i E t)$ during the propagation. 

We are now in a position to apply our algorithm to the calculation of 
time-dependent currents in one-dimensional model systems. The systems 
are initially in thermodynamic equilibrium. At time $t=0$, 
a time-dependent perturbation is switched on. In all the examples 
below the current is calculated according to Eq. (\ref{tdcoa}) 
\begin{eqnarray}
I(x,t) &=& 2 \int_{-k_F}^{k_F} 
\frac{\dr k}{2 \pi} \,
{\rm Im}\, \left( \psi_k^*(x,t) \frac{\dr}{\dr x} \psi_k(x,t) \right)
\nonumber \\ &=& 2 \int_{0}^{k_F} \frac{\dr k}{2 \pi} \,
{\rm Im}\, \left( \psi_k^* \frac{\dr}{\dr x} \psi_k +
\psi_{-k}^* \frac{\dr}{\dr x} \psi_{-k} \right)
\label{currdens}
\end{eqnarray}
where the prefactor $2$ comes from spin and $k_F=\sqrt{2 \ve_F}$ is the 
Fermi wavevector of a system with Fermi energy $\ve_{F}$. 

\subsection{DC bias}

As a first example we considered a system where the electrostatic potential 
vanishes identically both in the left and right leads as well as in the 
central region which is explicitly propagated. Initially, all single particle 
levels are occupied up to the Fermi energy $\ve_F$. At $t=0$ a 
constant bias is switched on in the leads and the time-evolution of the 
system is calculated. We chose the bias in the right lead as the negative 
of the bias in the left lead, $U_R = - U_L$.

The numerical parameters are as follows: the Fermi energy is 
$\ve_F=0.3$ a.u., the bias is $U_L=-U_R=0.05,\,0.15,\,0.25$ a.u., the central region 
extends from $x=-6$ to $x=+6$ a.u. with equidistant grid points with spacing 
$\Delta x=0.03$ a.u.. The $k$-integral in Eq.~(\ref{currdens}) is discretized 
with 100 $k$-points which amounts to a propagation 
of 200 states. The time step for the propagation was 
$\Delta t= 10^{-2}$ a.u.
\begin{figure}[htbp]
\begin{center}    
\includegraphics*[scale=0.5]{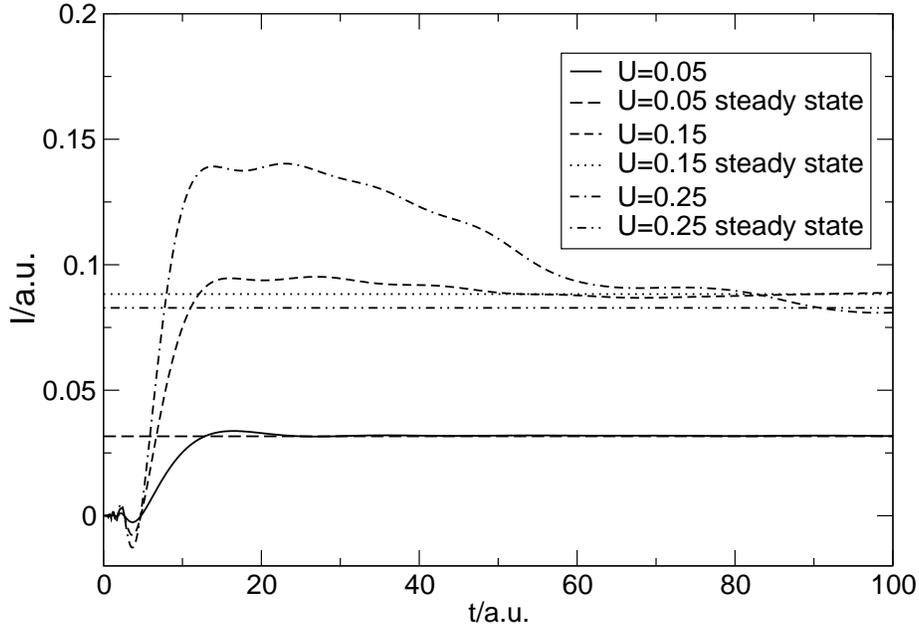}
\caption{ Time evolution of the current for a system where  
initially the potential is zero in the leads and the propagation 
region. At $t=0$, a constant bias with opposite sign in the left and right 
leads is switched on, $U=U_L=-U_R$ (values in atomic units). 
The propagation region extends from 
$x=-6$ to $x=+6$ a.u.. The Fermi energy of the initial state is $\ve_F=0.3$ 
a.u.. The current in the center of the propagation region is shown.}
\label{examp1}
\end{center}
\end{figure}

In Fig.~\ref{examp1} we have plotted the current densities at $x=0$ as a 
function of time for different values of the applied bias. 
As a first feature we notice that a steady state is achieved, 
in agreement with the discussion of Section \ref{steady}. 
The corresponding steady-state current $I^{(\rm S)}$ can be calculated from the 
Landauer formula. For the present geometry this leads to the steady current 
\begin{eqnarray}
I^{(\rm S)}=8e\int_{\max(U_{L},U_{R})}&&\frac{d\w}{2\p}[f(\w-U_{L})-f(\w-U_{R})]
\label{land} \\ &\times&
\frac{\sqrt{\w-U_{L}}\sqrt{\w-U_{R}}}
{\left[\sqrt{\w-U_{L}}+\sqrt{\w-U_{R}}\right]^{2}+U_{L}U_{R}
\left[
\frac{\mbox{\normalsize{$ \sin(l\sqrt{2\w}) $}} }{
\mbox{\normalsize{$ \sqrt{\w} $}} }
\right]^{2}},
\nonumber
\end{eqnarray}
where $l$ is the width of the central region. 
From Eq. (\ref{land}) with $l=12$ a.u. and $U_L=-U_R$, the 
numerical values for the steady-state currents  
are $0.0316$ a.u. ($U_{L}=0.05$ a.u.), $0.0883$ a.u. ($U_{L}=0.15$ a.u.) and 
$0.0828$ a.u. ($U_{L}=0.25$ a.u.). 
We see that our algorithm yields the same answers. Second, 
we notice that the onset of the current is delayed in relation to the 
switching time $t=0$. This is easily explained by the fact that the 
perturbation at $t=0$ happens in the leads only, {\em e.g.}, for  
$|x|> 6$ a.u., 
while we plot the current at $x=0$. In other words, we see the delay time 
needed for the perturbation to propagate from the leads to the center of 
our device region. We also note that the higher the bias the more the 
current overshoots its steady-state value for small times after switching 
on the bias. Finally it is worth mentioning that increasing the bias 
not necessarily leads to a larger steady-state current. 
\begin{figure}[htbp]
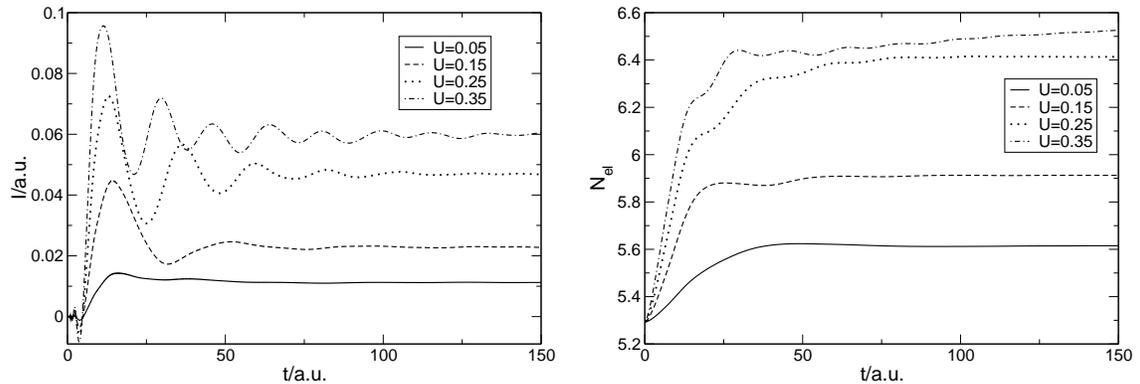

\begin{center}
\includegraphics*[scale=0.30]{fig6}$\quad$
\includegraphics*[scale=0.30]{fig7}\\[0.5cm]
\caption{Left panel: Time evolution of the current through a double 
square potential barrier in response to an applied constant bias (given in 
atomic units) in the left 
lead. The potential is given by $V(x) = 0.5$ a.u. for $5 \leq |x|\leq 6$ a.u. 
and zero otherwise, the propagation region extends from $x=-6$ to $x=+6$ a.u.. 
The Fermi energy of the initial state is $\ve_F=0.3$ a.u.. The current in the 
center of the structure is shown. Right panel: Time evolution of the total number of electrons 
in the region $|x|\leq 6$ for the same double square potential barrier. }
\label{examp2}
\end{center}
\end{figure}

In the second example we studied a double square potential barrier with 
electrostatic potential $V(x) = 0.5$ a.u. for $5$ a.u. $\leq |x| \leq 6$ a.u. 
and zero otherwise. This time we switch on a constant bias in the left lead 
only, {\em i.e.}, $U_R=0$. The Fermi energy for the initial state is 
$\ve_F=0.3$ a.u.. The central region extends from $x=-6$ to $x=+6$ a.u. with 
a lattice spacing of $\Delta x=0.03$ a.u.. Again, we use 100 different 
$k$-values to compute the current and a time step of 
$\Delta t = 10^{-2}$ a.u.. 

In Fig. \ref{examp2} (Left panel) we plot the current at 
$x=0$ as a function of time for several values of the applied bias $U=U_L$. 
Again, a steady state is achieved for all values of $U$. 
As discussed in Fig.~\ref{examp1} the transient current can exceed the
steady current; the higher the applied voltage the larger is this excess 
current and the shorter is the time when it reaches its maximum.
Furthermore, the oscillatory evolution towards the steady current solution
depends on the bias. For high bias the frequency of the transient 
oscillations increases. For small bias the electrons at the bottom of the  
band are not disturbed and the transient process is 
exponentially short. On the other hand, for a bias close to the Fermi 
energy the transient process decays as a power law, due to 
the band edge singularity. As pointed out in Section \ref{steady}, for 
non-interacting electrons the steady-state current develops by means 
of a pure dephasing mechanism. In our examples the transient process occurs 
in a femtosecond time-scale, which is much shorter than the  
relaxation time due to electron-phonon interactions. 

In Fig. \ref{examp2} (Right panel) we plot the time evolution of the total number of 
electrons in the device region for the same values of $U_L$. We see that as 
a result of the bias a quite substantial amount of charge is added to the 
device region. This result has important implications when simulating
the transport through an interacting system as the effective (dynamical) 
electronic screening is modified due not only to the external field but also
to the accumulation of charge state in the molecular device. This illustrates 
that linear response might not be an appropriate tool to tackle the dynamical
response and that we will need to resort to a full time-propagation
approach as the one presented in this review. 
Here we emphasize that all our calculations are done 
without taking into account the electron-electron interaction. If we had 
done a similar calculation with the interaction incorporated in a 
time-dependent Hartree or time-dependent DFT framework we would expect the 
amount of excess charge to be reduced significantly as compared to 
Fig. \ref{examp2}.

\subsection{Time-dependent biases}

In the previous Section we have shown how a steady current 
develops after the switching on of a constant bias and discussed 
the transient regime. Here we exploit the versatility of our proposed  
algorithm for studying different kinds of time-dependent 
biases.

\begin{figure}[htbp]
\begin{center}
\includegraphics*[scale=0.5]{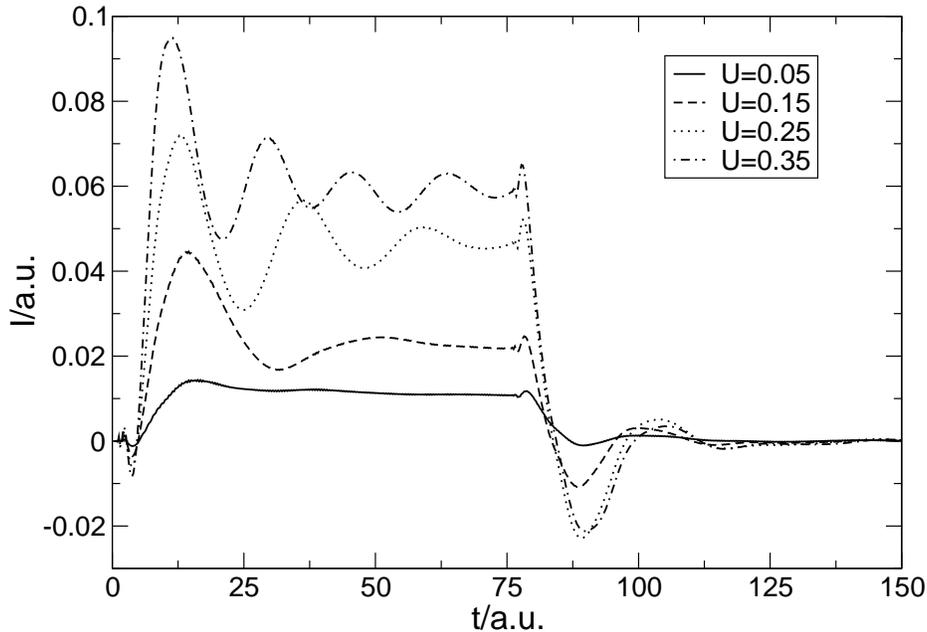}
\caption{Same system of Fig. \ref{examp2} exposed to a suddenly 
switched on bias at $t=0$. The bias is then turned off at $t=75$ a.u.
The current is measured in the middle of the central region.}
\label{onoff} 
\end{center}
\end{figure}

\begin{figure}[htbp]
\begin{center}
\includegraphics*[scale=0.5]{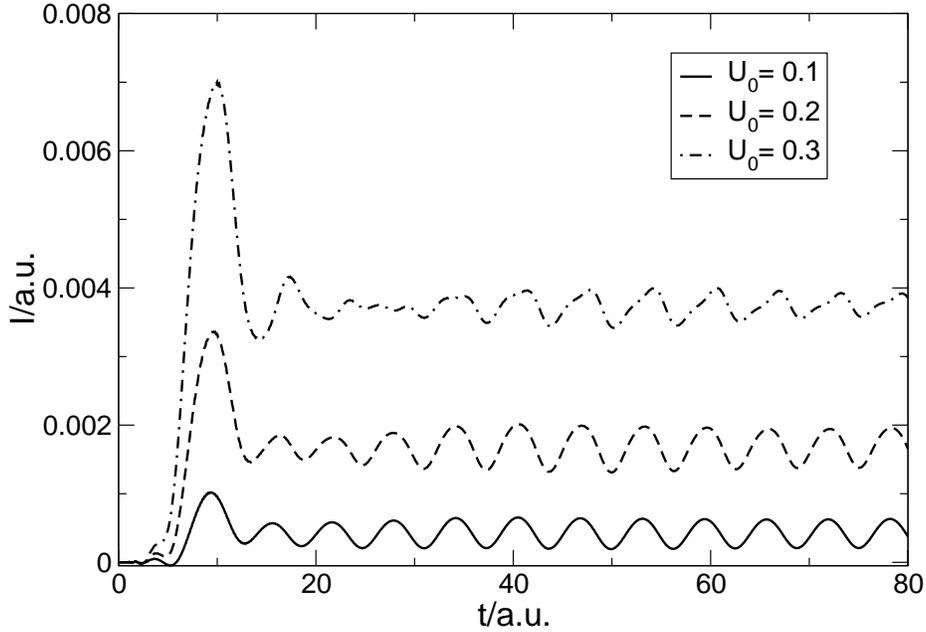}
\caption{Time evolution of the current for a square potential 
barrier in response to a time-dependent, harmonic bias in the left lead, 
$U_L(t) = U_0 \sin(\omega t)$ for different amplitudes $U_0$ (values in a.u.) 
and frequency $\omega=1.0$ a.u.. The potential is given by $V(x)=0.6$ a.u. for 
$|x|\leq 6.0$ a.u. and zero otherwise. The propagation region extends from 
$x=-6$ to $x=+6$ a.u.. The Fermi energy of the initial state is 
$\ve_F=0.5$ a.u.. The current at $x=0$ is shown. }
\label{examp3} 
\end{center}
\end{figure}

As a first example we consider how the 
current responds to a sudden switching off of the bias. For 
comparison we have considered the same double square potential barrier 
of Fig. \ref{examp2} subject to the same suddenly switched on bias, 
but we have turned off the bias at $t=75$ a.u. The results (obtained 
with the same parameters of Fig. \ref{examp2}) are 
displayed in Fig. \ref{onoff}. We observe that the current shows a 
rather well pronounced peak shortly after switching off the 
perturbation. The amplitude of the peak is proportional to the 
originally applied bias.
This peak always overshoots the value of the current at the steady 
state. Another interesting feature is the fact that after turning off the 
bias the transient currents show only two oscillations around the zero 
current limit and the transient time for switching off is much shorter than 
for switching on a high bias. 

We have also addressed the simulation of AC-transport.
We computed the current for a single square potential 
barrier with $V(x) = 0.6$ for $|x|<6$ and zero otherwise. Here we 
applied a time-dependent bias of the form $U_L(t) = U_{0} \sin(\omega t)$ 
to the left lead. The right lead remains on zero bias. The numerical 
parameters are: Fermi energy $\ve_F=0.5$ a.u., device region from $x=-6$ to 
$x=+6$ a.u. with lattice spacing $\Delta x=0.03$ a.u.. The number of 
$k$-points is 100 and the time step is $\Delta t=  10^{-2}$ a.u..  
In Fig.~\ref{examp3} we plot the current at $x=0$ as a function of time for 
different values of the parameter $U_{0}=0.1,0.2,0.3$ a.u. The frequency was 
chosen as $\omega=1.0$ a.u. in both cases. Again, as for the DC-calculation  
discussed above, we get a transient that overshoots the average current 
flowing through the constriction; again, this excess current is larger the 
higher the applied voltage. Also, after the transient we obtain a current 
through the system with the same period as the applied bias. Note, however, 
that (especially for the large bias), the current is not a simple harmonic 
as the applied AC bias.

\begin{figure}[htbp]
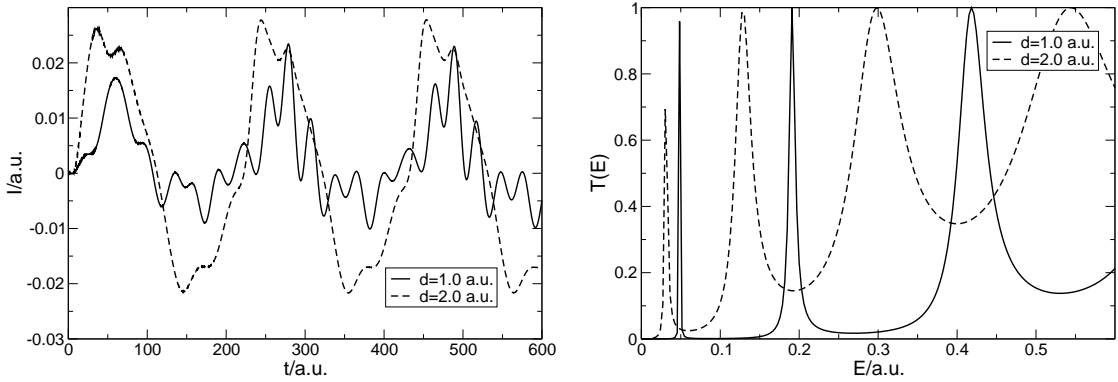

\begin{center}
\includegraphics*[scale=0.30]{curr}$\quad$
\includegraphics*[scale=0.30]{transmission}\\[0.5cm]
\caption{Left panel: Time evolution of the current for a symmetric 
double square potential 
barrier in response to a time-dependent, harmonic bias in the left lead, 
$U_L(t) = U_0 \sin(\omega t)$ with $U_{0}=0.15$ a.u. and $\w=0.03$ a.u.
for different thickness $d=1$ and $d=2$ a.u. of the barriers. 
Right Panel: Transmission function of the same double square potential 
barrier for $d=1$ and $d=2$ a.u.}
\label{transcurr}
\end{center}
\end{figure}

Exposing the system to an AC bias also allows us to acquire 
information about the excitation energies of the  molecular device.
In Fig. \ref{transcurr} (Left panel) we plot the time dependent 
current for a symmetric double square potential barrier in response to a 
harmonic bias in the left lead, $U_L(t) = U_0 \sin(\omega t)$, with 
$U_{0}=0.15$ a.u. and $\w=0.03$ a.u.. The Fermi energy of the initial state is 
$\ve_F=0.3$ a.u. and the current at $x=0$ is shown. 
The central region extends from $x=-6$ to $x=6$ a.u. 
with lattice spacing $\Delta x=0.03$ a.u.
and the potential $V(x)$
in region $C$ is given by $V(x)=0$ for $|x|/{\rm a.u.}<(6-d)$  and 
$V(x)=0.5$ a.u. for $(6-d)<|x|/{\rm a.u.}<6$. 
The number of $k$-points is 100 and the time step is $\Delta t=  10^{-2}$ a.u.. 
We have studied barriers of 
different thickness $d=1$ a.u. and $d=2$ a.u.. For $d=2$ a.u.
we observe small oscillations superimposed to the oscillations of 
frequency $\w=0.03$ a.u. driven by the external AC field. Such small 
oscillations have frequency $\simeq 0.23$ and can be understood by 
looking at the transmission function $T(E)$ in the Right panel of 
Fig. \ref{transcurr}. For $d=2$ a.u. both the second and third 
peaks of $T(E)$ are in the energy window 
$(\ve_{F}-U_{0},\ve_{F}+U_{0})=(0.15,0.45)$ a.u..
The energy difference between these two peaks corresponds to a good 
extent to the frequency of the superimposed oscillations.
On the contrary, for $d=1$ a.u. only one peak of the transmission function 
$T(E)$ is contained in the aforementioned energy window 
and no superimposed oscillations are clearly visible. This example 
shows the AC quantum transport can be used also for probing 
molecular devices.

\subsection{History dependence}

\begin{figure}[htbp]
\begin{center}
\includegraphics*[scale=0.5]{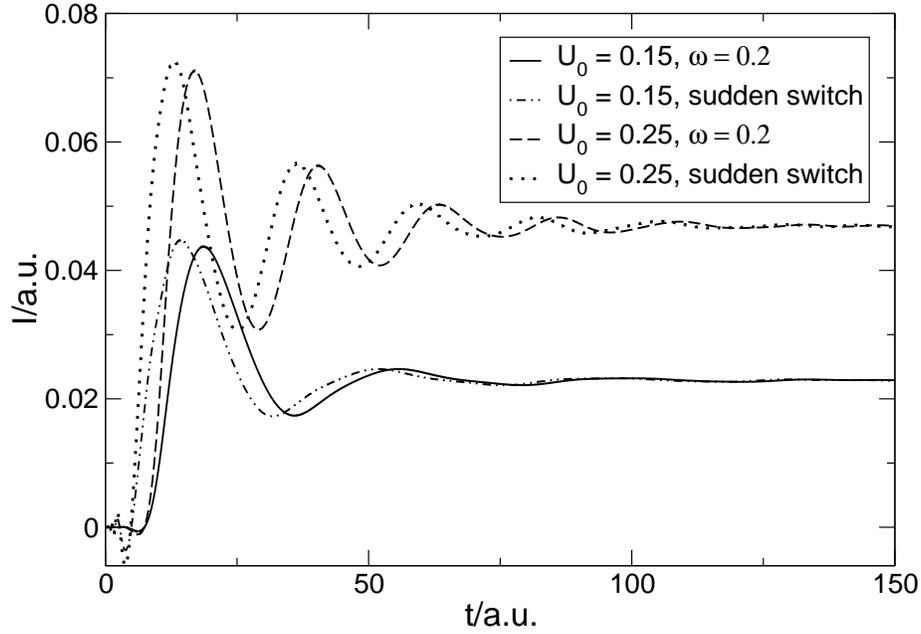}
\caption{ Time evolution of the current for a double 
square potential barrier when the bias is switched on in two different 
manners: in one case, the bias $U_0$ is suddenly switched on at $t=0$ 
while in the other case the same bias is achieved with a smooth switching
$U(t) = U_0 \sin^2(\omega t)$ for $0<t<\pi/(2 \omega)$. The parameters 
for the double barrier and the other numerical parameters are the same as 
the ones used in Fig.~\ref{examp2}. $U_0$ and $\omega$ given in atomic units.}
\label{examp2a}
\end{center}
\end{figure}

In Fig.~\ref{examp2a} we show time-dependent currents for the same double 
barrier as in Fig.~\ref{examp2} for two different ways of applying the 
bias in the left lead: in one case the constant bias $U_0$ is switched 
on suddenly at $t=0$ (as in Fig.~\ref{examp2}), in the other case the 
constant $U_0$ is achieved with a smooth switching $U(t)=U_0 
\sin^2(\omega t)$ for $0<t<\pi/(2 \omega)$. As expected and in agreement 
with the results of Section \ref{steady}, the same steady state 
is achieved after the initial transient time. However, the transient 
current clearly depends on how the bias is switched on. 

\begin{figure}[htbp]
\begin{center}
\includegraphics*[scale=0.5]{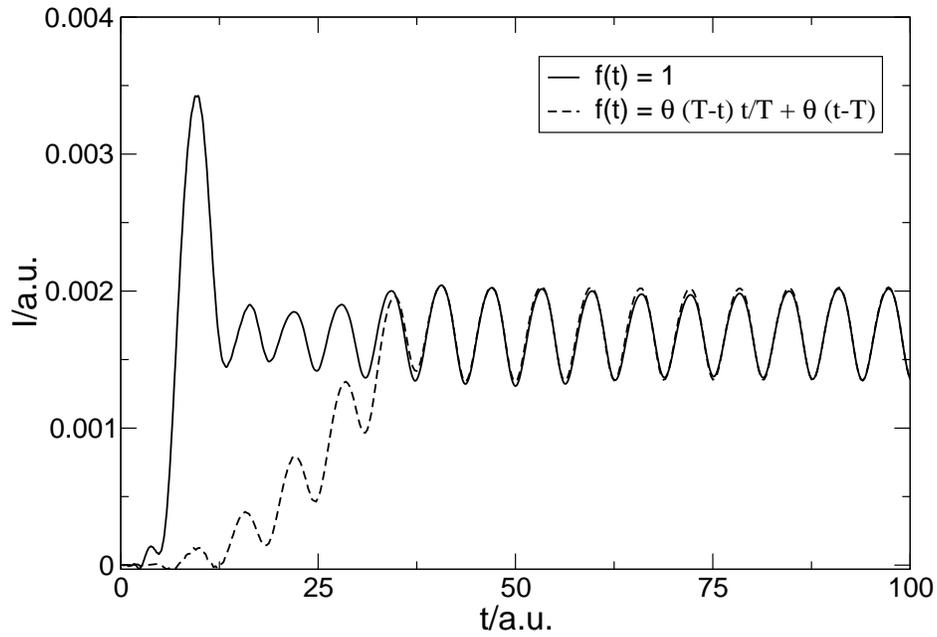}
\caption{Time evolution of the current for a square potential 
barrier in response to a time-dependent, harmonic bias in the left lead, 
$U_L(t) = U_0 f(t)\sin(\omega t)$ with $U_0=0.2$ a.u.
and frequency $\omega=1.0$ a.u.. The system and the parameters used 
are the same as in Fig. \ref{examp3}. The current at $x=0$ is shown
for two different ``switching on'' functions $f(t)$.}
\label{history}
\end{center}
\end{figure}

According to the result in Eq. (\ref{qaasym}), for noninteracting electrons 
the independence of the history is not limited to 
steady-state regimes. The long-time behaviour of currents $I(t)$ 
and $I'(t)$ induced by biases $U_{\a}(t)$ and $U'_{\a}(t)$ 
does not change provided $U_{\a}-U'_{\a}\ra 0$ for $t\ra \inf$. 
For instance, the current response to an AC bias has the same periodic 
modulation and the {\em same phase} independently of how the AC bias is 
switched on.
In Fig. \ref{history} we plot the time-dependent current for the same 
system (and using the same parameters) of Fig. \ref{examp3}. The bias 
remains on zero in the right lead. In the left lead we applied a 
time-dependent bias of the form $U_{L}(t)=U_{0}f(t)\sin(\w t)$, with 
$U_{0}=0.2$ a.u., $\w=1.0$ a.u., and we considered two different 
``switching on'' functions $f(t)$. The first is $f(t)=1$ (as in 
Fig. \ref{examp3}) while the second is a ramp-like switching-on 
$f(t)=\theta(T-t)t/T+\theta(t-T)$ with $T= 30$ a.u.. As expected, and in 
agreement with Eq. (\ref{qaasym}), the current has the same behaviour 
in the long-time limit.

\subsection{Pumping current: preliminary results}
\label{pumping}

Our algorithm is also well-suited to study pumping of electrons. An 
electron pump is a device which generates a DC current between two 
electrodes kept at the same bias. The pumping is achieved by applying 
a periodic gate voltage depending on two or more parameters. Electron 
pumps have been realized experimentally, e.g., for an open semiconductor 
quantum dot \cite{SwitkesMarcusCampmanGossard:99} where pumping was achieved 
by applying two harmonic gate voltages with a phase shift. 

In the literature, different techniques have been used to discuss electron 
pumping theoretically. Brouwer \cite{Brouwer:98} suggested a scattering 
approach to describe pumping of non-interacting electrons which has been used, 
e.g., to study pumping through a double barrier \cite{WeiWangGuo:00}. 
Nonequilibrium Green's function techniques have been used to study 
pumping in tight-binding models of coupled quantum dots 
\cite{StaffordWingreen:96}. Alternatively, Floquet theory which describes 
evolution of a quantum system under the influence of time-periodic fields 
is also well-suited to describe pumping 
\cite{KohlerLehmannHaenggi:05}.

\begin{figure}[htbp]
\begin{center}
\includegraphics*[scale=0.45]{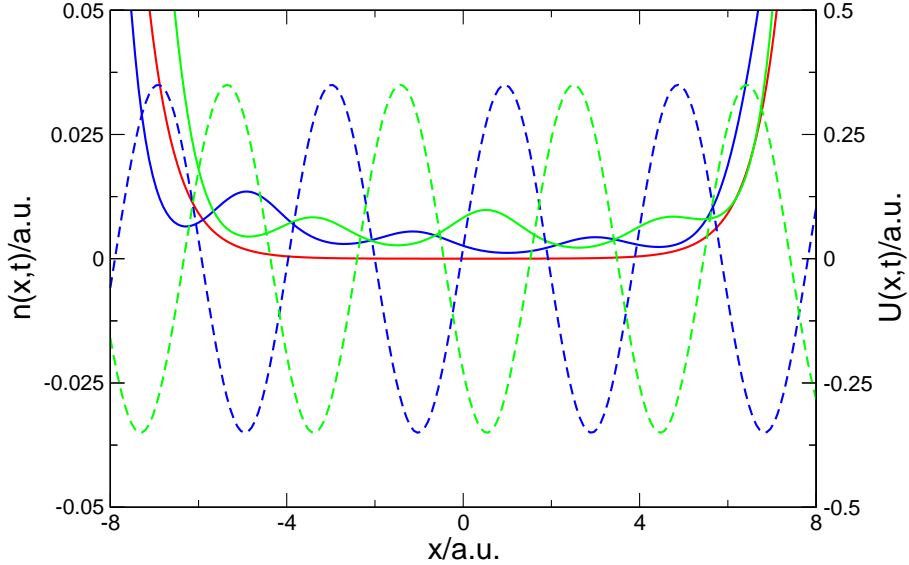}
\caption{Snapshots of the density for and the travelling potential wave at 
various times for pumping through a single square barrier by a travelling 
wave. The barrier with 
height $0.5$ a.u. extends throughout the propagation window from 
$x=-8$ to $x=+8$. The leads are on zero potential and the Fermi level is 
at $0.3$ a.u.. The travelling potential wave is restricted to the propagation 
window $|x|< 8$ and has the form $U(t) = U_0 \sin(q x - \omega t)$ with 
amplitude $U_0=0.35$ a.u., wave number $q=1.6$ a.u. and frequency 
$\omega=0.2$ a.u.. The initial density is given by the red line.}
\label{pump1}
\end{center}
\end{figure}

\begin{figure}[htbp]
\begin{center}
\includegraphics*[scale=0.7,angle=-90]{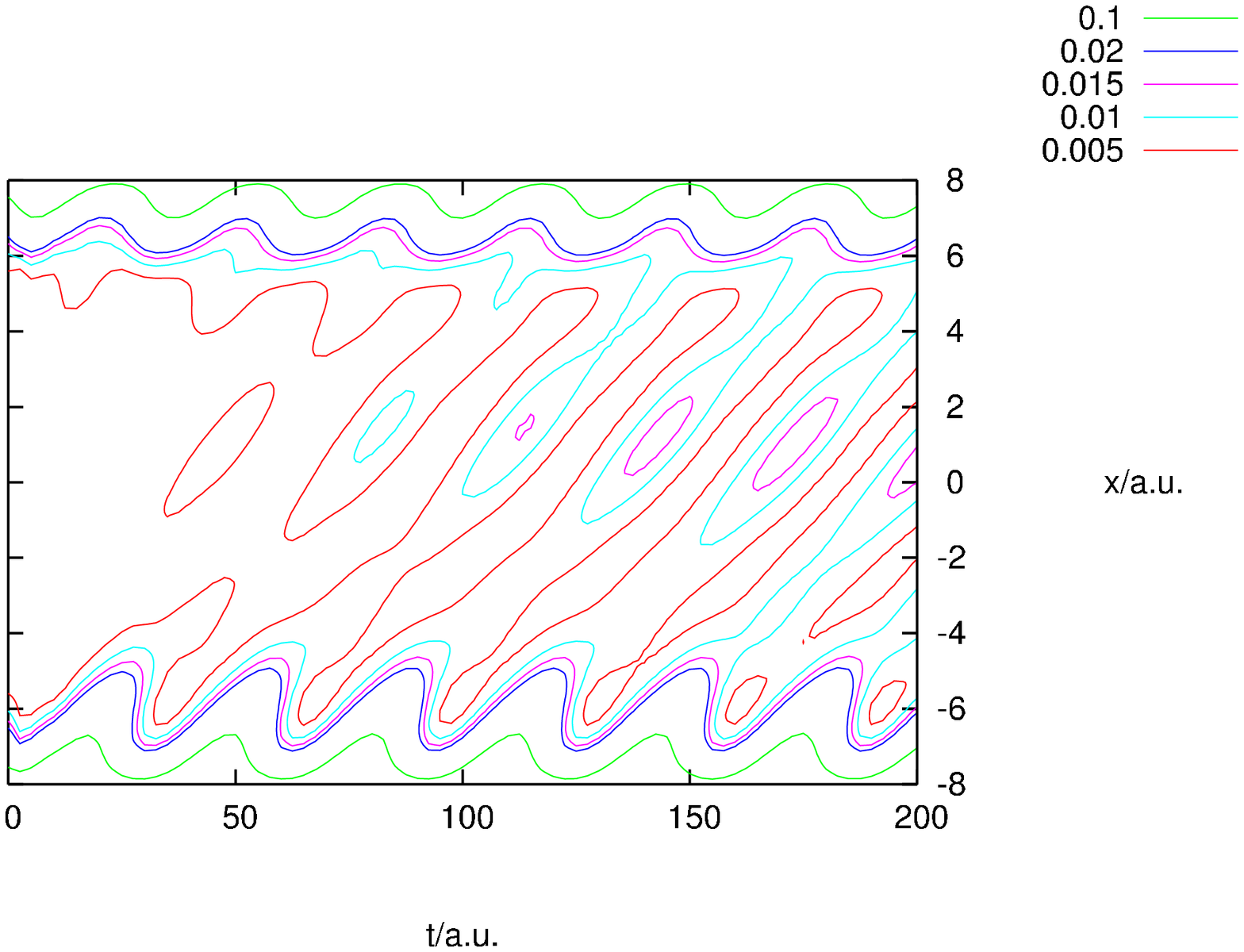}
\caption{Contour plot of the time-dependent density for pumping through a 
single square barrier by a travelling potential wave. The parameters are the 
same as for Fig.~\protect\ref{pump1}.}
\label{pump2}
\end{center}
\end{figure}

As a first example of electron pumping we have calculated the time evolution 
of the density for a single square barrier exposed to a travelling potential 
wave $U(t) = U_0 \sin(q x - \omega t)$. The wave is spatially restricted to 
the explicitly treated device region which in our case also coincides with 
the static potential barrier. Some snapshots of the density and the 
potential wave are shown in Fig.~\ref{pump1}. The density in the device region 
clearly exhibits local maxima in the potential minima and is transported 
in pockets by the wave. This is also evident in Fig.~\ref{pump2} where we 
show the time-dependent density as function of both position and time 
throughout the propagation. The density contour lines show transport of 
electrons from the left lead at $x=-8$ to the right lead at $x=+8$ a.u.. The 
pumping mechanism in this example resembles pumping of water with the 
Archimedean screw. 

\begin{figure}[htbp]
\begin{center}
\includegraphics*[scale=0.45]{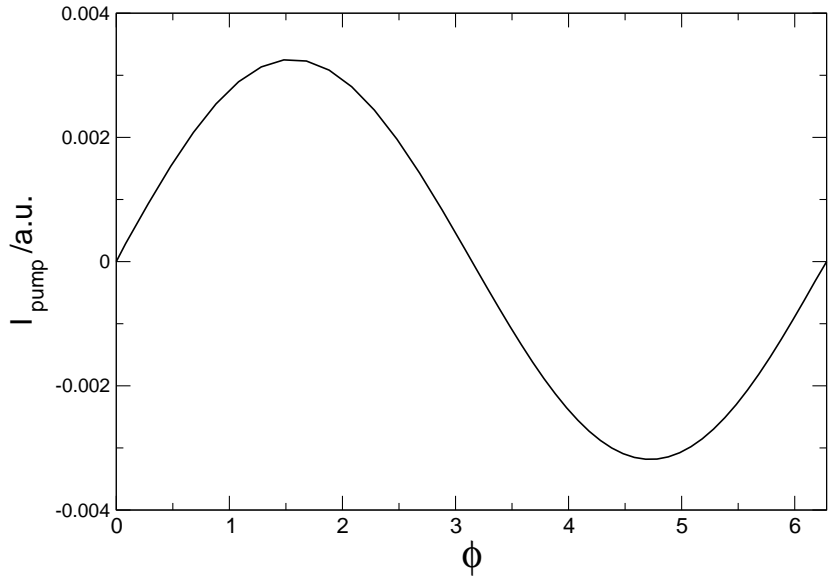}
\caption{Parametric pumping through a double square barrier. The device region 
extends from $x=-6$ to $x=+6$ a.u., the static potential has the value 
$0.525$ a.u. for $5<|x|<6$ a.u. and zero elsewhere in the device. Pumping 
is achieved by harmonic variation of the barriers, i.e., 
$U(x,t) = U_0 \sin(\omega t)$ for the left barrier ($-6$ a.u.$ <x<-5$ a.u.) and 
$U(x,t) = U_0 \sin(\omega t + \phi)$ for the right barrier 
($5$ a.u.$ <x<6$ a.u.). The DC component of the pump current is displayed as 
a function of the phase $\phi$. The parameters are: $U_0=0.25$ a.u., 
$\omega = 0.25$ a.u. and the Fermi energy is $\varepsilon_F=0.5$ a.u..}
\label{pump3}
\end{center}
\end{figure}

As a second example we have calculated pumping through a double square barrier 
by applying two harmonic gate voltages with a phase difference to the barrier 
potentials, i.e., $U(x,t) = U_0 \sin(\omega t)$ for the left barrier and 
$U(x,t) = U_0 \sin(\omega t + \phi )$ for the right barrier. Fig.~\ref{pump3} 
shows the DC component of the pump current as a function of the phase $\phi$ 
which has a sinusoidal dependence for our parameter values. This is in 
agreement with similar results of Ref.~\onlinecite{WeiWangGuo:00} for small 
amplitudes of the AC bias which were obtained using Brouwer's approach. In 
addition, this example may be interpreted as a very simple model to describe 
the experiment of Ref.~\onlinecite{SwitkesMarcusCampmanGossard:99}. 

\section{Conclusions and perspectives}
\label{conclusion}

In this review we have given a self-contained introduction to our 
recent approach to quantum transport. In essence our approach combines two 
well-established theories for the description of non-equilibrium phenomena 
of interacting many-electron systems. 

On the one hand there is the formalism of non-equilibrium Keldysh-Green 
functions. Although this approach in principle can be used to study 
interaction effects, here we only used it in the context of non-interacting 
electrons. The reason for this is that the self-energy of interacting electrons
(which is not to be confused with the embedding self-energy) is long-range and 
nonlocal. In our scheme which partitions space in left and right leads as well 
as the central device region, this non-locality is extremely difficult to deal 
with in a rigorous manner. 

On the other hand, the NEGF formalism for (effectively) non-interacting 
electrons can easily be combined with the second approach for time-dependent 
many-particle systems, namely time-dependent density functional theory. Just 
as the NEGF formalism, TDDFT in principle gives the correct time-dependent 
density of the interacting system (if the exact exchange-correlation potential 
is used). Moreover, the time-dependent effective single-particle potential 
of TDDFT is a local and multiplicative potential which is crucial for 
practical use within the partitioning scheme for transport.

In combining the NEGF and TDDFT approaches we have presented a formally
rigorous approach towards the description of charge transport using
an open-boundary scheme within TDDFT. We have implemented a
specific time-propagation scheme that incorporates transparent
boundaries at  the device/lead interface in a natural way.
In order to have a clear
definition of a device region in Fig.~\ref{system} we assumed
that an applied bias can be described by adding a spatially constant
potential shift in the macroscopic part of the leads. This implies an
effective ``metallic screening'' of the constriction. The screening limits
the spatial extent of the induced density created by
the bias potential or the external field
to the central region. 
Our time-dependent scheme allows to extract both AC and DC I/V
device characteristics and it is ideally suited to describe
external field (photon) assisted processes.

In order to illustrate the performance and potential of the method we have 
implemented it for one-dimensional model systems and applied it to a 
variety of transport situations: we have shown that a steady-state current 
is always reached upon application of a DC bias. For a harmonic AC bias, the 
resulting AC current need not be harmonic. 
In the case of systems at DC bias without any source of dissipation it
is known that the steady-current is
independent of the history of the process\cite{saprb04}.
We have explicitly demonstrated this history independence for two different
switching processes of the external bias. The history independence for 
non-interacting electrons not only applies for DC but also for AC bias which 
we have also demonstrated in a numerical example. Since we can compute current 
densities locally, we are not restricted to currents deep inside the leads. 
In one example we have analyzed the time evolution of the density for localized 
states which are only weakly coupled to the reservoirs. Finally, we have shown 
a few simple applications of our algorithm to electron pumping. 

The list of the example calculations presented here already demonstrates the 
versatility and flexibility of our algorithm. It includes the Landauer 
formalism as the long-time limit for systems under DC bias and allows to 
study transients. Moreover, it can deal with periodic time-dependent fields 
(which are usually treated with the Floquet formalism) but is applicable to 
nonperiodic conditions as well \cite{OriolsAlarconFernandezDiaz:05}. 

Most theoretical approaches to transport adopt open boundary conditions and 
assume that transport is ballistic,{\em i.e.}, under steady state conditions 
inelastic collisions are absent and dissipation occurs only in the idealized 
macroscopic reservoirs. This might be an unrealistic assumption for transport 
through single molecules, in particular when the device is not operated 
in the regime of small bias and linear response. When inelastic scattering 
dominates this picture is not applicable.
In particular, experiments \cite{kushmerick,smit,dekker} indicate that 
inelastic
scattering with lattice vibrations is present at sufficiently large bias,
causing local heating of contacts and molecular devices.
In addition, current-induced forces might lead to bond-breaking and 
electromigrations.

In a joint collaboration with Verdozzi and Almbladh, 
one of us has included the nuclear degrees of freedom at a classical 
level\cite{vsa2006}. The initial ground state (consisting of bound, 
resonant and scattering states) has been calculated 
self-consistently. Also, the time-propagation 
algorithm of Section \ref{kst} has been generalized 
to evolve the system electrons+nuclei in the Ehrenfest approximation.
Several aspects of the electron-ion interaction in quantum 
transport have been investigated.
 
Electron correlations are also important in molecular conductors,
for example, Coulomb blockade effects dominate the transport in quantum dots.
Short-range electron correlations seems to be relevant in order to get
quantitative description of I/V characteristics in molecular
constrictions\cite{delaney:04,everts:04,ferreti}.
In particular it is commonly assumed that the energy scales for
electron-electron and electron-phonon interactions are different and
could be treated separately. However, the metallic screening of the
electrodes considerably reduces the Coulomb-charging energy (from eV to
meV scale). In this regime the energy scale for the two
interactions merge and they need to be treated on the same footing. 
We would like to emphasize that our scheme allows for a consistent treatment 
of electronic and ionic degrees of freedom.                     

It is clear that the quality of the TDDFT functionals is of crucial
importance. In particular, exchange and correlation functionals for
the non-equilibrium situation are required.
Time-dependent linear response theory for DC-steady state
has been implemented in Ref.~\onlinecite{baer:04}
within TDLDA assuming jellium-like electrodes (mimicked by complex
absorbing/emitting potentials). It has been shown that the
DC-conductance changes considerably from the standard Landauer value.
Therefore, a systematic study of the TDDFT
functionals themselves is needed. A step beyond
standard adiabatic approximations and exchange-only potentials
is to resort to many-body schemes based on 
perturbative expansions\cite{exc,punk}, iterative schemes\cite{lucia}, or 
variational-functional formulations\cite{vbdvls2005}. 
Another path is to explore exchange-correlation functionals
that depend implicitly \cite{dbg1997,kb2004} or
explicitly \cite{vignale,tokatly} on the current density.

\section*{Acknowledgments}
This work was supported in part by the Deutsche Forschungsgemeinschaft, 
by the EU Research and Training Network EXCITING and by the NANOQUANTA
Network of Excellence.

\end{document}